\documentclass[aps,twocolumn,showpacs,email,amssymb,nobibnotes]{revtex4}

\usepackage{bm}
\usepackage{graphicx}
\usepackage{amsmath}
\usepackage{amsthm}
\usepackage{eucal}
\usepackage{amssymb}
\usepackage{mathrsfs}
\usepackage{enumerate}
\usepackage{xcolor}

\newcommand{\eref}[1]{Eq.~(\ref{#1})}

\newcommand{\fref}[1]{Fig. \ref{#1}}

\newcommand{\sref}[1]{section \ref{#1}}

\newcommand{\Aref}[1]{Appendix \ref{#1}}
\newcommand{\comments}[1]{}

\begin{document}

\title{Stationary solutions for metapopulation Moran models with mutation and selection}

\author{George W.~A.~Constable$^{1,2}$ and Alan J.~McKane$^{2,3}$}

\affiliation{$^{1}$Department of Ecology and Evolutionary Biology,
Princeton University, Princeton NJ 08544-2016, U.S.A. \\
$^{2}$Theoretical Physics Division, 
School of Physics and Astronomy,
The University of Manchester, Manchester M13 9PL, United Kingdom \\
$^{3}$Isaac Newton Institute, 20 Clarkson Road, Cambridge CB3 0EH, 
United Kingdom}

\begin{abstract}
We construct an individual-based metapopulation model of population genetics featuring migration, mutation, selection and genetic drift. In the case of a single `island', the model reduces to the Moran model. Using the diffusion approximation and timescale separation arguments, an effective one-variable description of the model is developed. The effective description bears similarities to the well-mixed Moran model with effective parameters which depend on the network structure and island sizes, and is amenable to analysis. Predictions from the reduced theory match the results from stochastic simulations across a range of parameters. The nature of the fast-variable elimination technique we adopt is further studied by applying it to a linear system, where it provides a precise description of the slow-dynamics in the limit of large timescale separation.  
\end{abstract}
\pacs{87.10.Mn, 05.40.-a, 02.50.Ey}

\maketitle

\section{Introduction}\label{secIntroduction}

In the theory of population genetics, the central aim is to understand the change in allele frequencies in populations which are subject to the processes of mutation, genetic drift, natural selection, and migration between subpopulations~\cite{roughgarden_1979,halliburton_2004,hartl_1989}. Even in the simple case of a gene with two alleles at a single locus in a haploid individual, this is still a challenging task. In two recent papers~\cite{constable_phys,constable_bio} we introduced an approximation procedure which reduced a model of migration between an arbitrary number of islands to one on a single island, but with effective parameters which included those describing the network structure and the population size of the islands. The analytic predictions of the reduced model were generally found to be in very good agreement with simulations of the original model in the case when genetic drift and selection were included. 

The aim of this paper is twofold. One is to include mutation in the formalism; our original studies did not include mutation, only drift, migration and selection. Adding mutation changes the long-time nature of the system. Without mutation, one of the alleles eventually dies out (the other is then fixed), and the quantities that one attempts to predict are the probabilities of this happening for the various alleles and how long this takes on average. With mutation present, they need never die out, and instead one asks what the stationary probability distribution function (pdf) of the system is. The other aim of the paper is to explore the nature of the approximation further. In our previous papers~\cite{constable_phys,constable_bio} we stressed the intuitive understanding of the approximation, but did not pursue the questions of why the approximation worked quite so well, and how and why it was superior to another scheme we used in the past~\cite{constable2013}. We address these questions here.

To carry out the approximation one begins with a neutral metapopulation Moran model. It is neutral because the birth/death rates of both alleles are the same. It is a metapopulation model because it consists of an arbitrary number of islands (or demes in the language of population genetics) between which genes can migrate. The islands are labeled by an index $i=1,\ldots,\mathcal{D}$ and the migration rate from island $j$ to island $i$, denoted by $m_{ij}$, is assumed to be given. Finally, the evolutionary dynamics is taken to be a Moran process~\cite{ewens_2004}, since this is perhaps the simplest model to analyze which has the necessary structure to illustrate the approach. The effect of selection (and, as we shall discuss, mutation) is governed by a small parameter, and can be added on to the neutral metapopulation model as a perturbative correction.

\begin{figure}[t]
\includegraphics[width=0.44\textwidth]{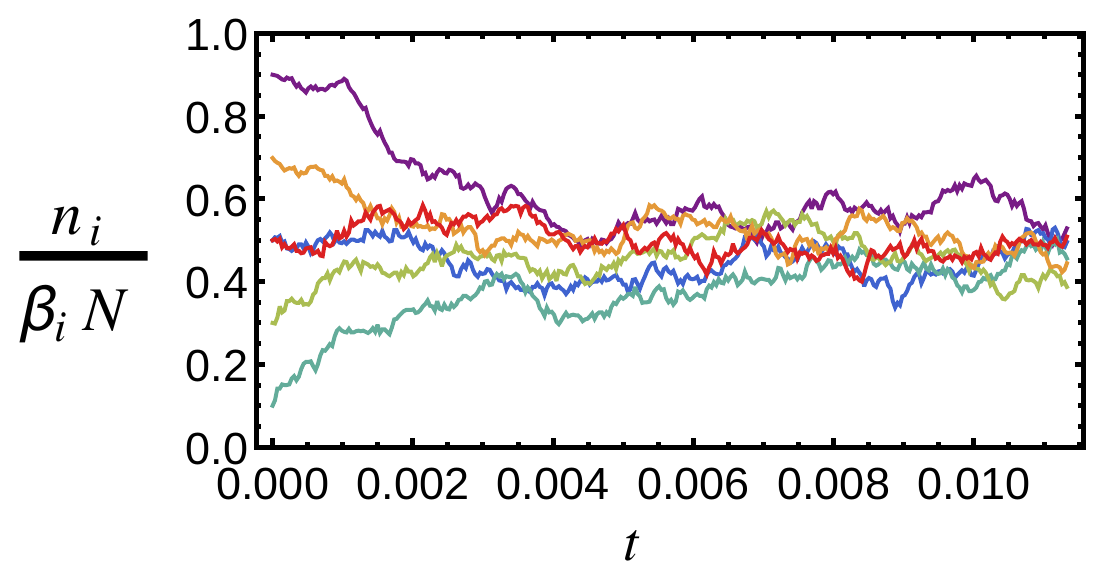}
\includegraphics[width=0.44\textwidth]{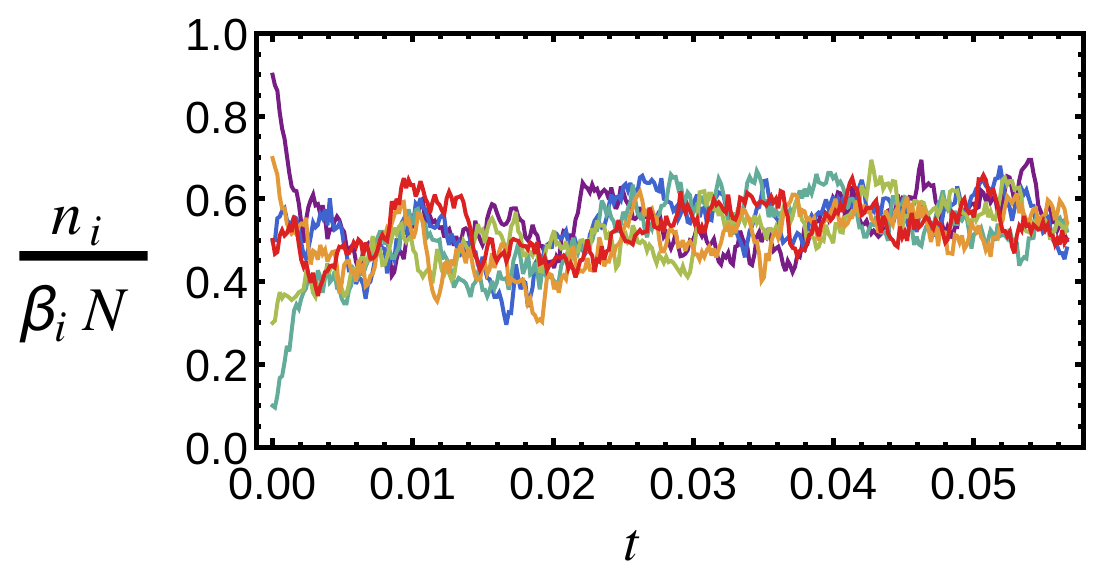}
\caption{(Color online) This figure shows the fractions of type $X$ individuals on each island $i$, $n_{i}/\beta_{i}N$, for a neutral six-island metapopulation Moran model. Over short times (upper panel) the proportions of individuals on each island become almost equal. After this the trajectories of each island are coupled, so that over long times (lower panel) the system behaves approximately as a well-mixed model with effective parameters. }
\label{fig_timescale_separation}.
\end{figure}

The basic idea behind the approximation is that, although the neutral metapopulation Moran model is described by $\mathcal{D}$ variables, after a short time (relative to those that interest us here) the dynamics of the system can be described by a single variable. This is illustrated in \fref{fig_timescale_separation}, where it is seen that effectively the composition of the islands cease to differ to any great extent after a short time. In mathematical terms there are $\mathcal{D}-1$ fast variables which quickly collapse onto a center manifold; the system effectively then moves on this (slow) center manifold until eventually one of the alleles becomes extinct. Including selection or mutation on top of this structure changes some of the details, but as long as the selection strength or the mutation rates are small, there will still be a timescale separation between the fast variables and the single slow variable, and the same methodology will be applicable. 

The outline of the paper is as follows. In \sref{sec_model} the general model, which includes all four evolutionary processes, is formulated, and the diffusion approximation is applied to obtain the Fokker-Planck equation (FPE) for the stochastic dynamics of these processes. Restricting our attention to the case in which selection is not present, a reduced-dimension description of the metapopulation model with mutation is obtained in \sref{sec_mutation}. We find that the reduced model is exactly a one-island Moran model with an effective noise-strength and effective mutation rates which depend on the migration and mutation rates, and island sizes of the original model. In \sref{sec_mutation_selection}, the reduced system is derived with selection included. In all cases we show that the approximation leads to results which agree extremely well with simulations of the original model. The approximation method we use is one of a large class of similar methods, and in \sref{sec_relate} we look at a simple model to obtain some insight into the features which make it successful. Finally, in \sref{sec_conclusion}, we conclude.

\section{Formulation of the metapopulation Moran model with mutation and selection}\label{sec_model}

For the purposes of this paper, it is most clear to introduce the model in its full generality, that is, with mutation and selection included. This is because it is essentially no more complicated to make the diffusion approximation with all processes included as it is to use just the neutral version of the model (see \Aref{app_diffusion} and also Appendix A of \cite{constable_bio}). 

We will assume the simplest constituents of the Moran model: only two types of haploid organisms, one carrying allele X and the other allele Y. Each of the $\mathcal{D}$ islands contains $n_i$ organisms of type $X$ and $N_i - n_i$ organisms of type $Y$, where the number of organisms on island $i$, $N_i$, is fixed. The state of the system at any given time is then given by the vector $\bm{n}=(n_1,\ldots,n_{\mathcal{D}})$. The islands may contain different numbers of organisms, but it is assumed that these are not orders of magnitude different from each other, so that we may write $N_i = \beta_i N$, where $N$ is a typical number of organisms on an island, and $\beta_i$ is a number of order one.

The stochastic dynamics is assumed to be Markovian, that is, the rate of making a transition from the current state $\bm{n}$ to a new state $\bm{n'}$ only depends on these two states, and not on the previous history of the system. We denote this rate by $T(\bm{n'}|\bm{n})$. The dynamics is then completely specified by the master equation~\cite{van_Kampen_2007}
\begin{equation}
\frac{d p(\bm{n},t)}{d t} = \sum_{\bm{n}'\neq\bm{n}}\,\left[ T(\bm{n}|\bm{n}')
p(\bm{n}',t) -  T(\bm{n}'|\bm{n})p(\bm{n},t) \right],
\label{master}
\end{equation}
with given initial conditions. The precise form of the transition rates $T(\bm{n'}|\bm{n})$ define the model. 

The transition rates for the neutral metapopulation Moran model are a natural generalization of the rules of the well-mixed (one-island) Moran model~\cite{moran_1962}. They have the form~\cite{constable_phys,constable_bio} 
\begin{eqnarray}\label{eq_neutral_rates}
 T_{0}( n_{i}+1 | n_{i} ) &=& \sum_{j=1}^{\mathcal{D}} \left( f_{j} \right) \left(\frac{n_{j}}{\beta_{j} N} \right) \left( m_{ij} \right) \left(\frac{\beta_i N - n_{i}}{\beta_{i}N - \delta_{ij}} \right),
\nonumber \\			
 T_{0}(n_{i}-1 | n_{i} ) &=& \sum_{j=1}^{\mathcal{D}} \left( f_{j} \right) \left(\frac{\beta_{j}N-n_{j}}{\beta_{j}N} \right) \left( m_{ij} \right) \left(\frac{n_{i}}{\beta_{i}N - \delta_{ij}} \right), \nonumber \\ \nonumber \\
\end{eqnarray}
where the dependence of the probability transition rates, $T_{0}(\bm{n'}|\bm{n})$, on elements of $\bm{n}$ that do not change in the transition has been suppressed. The subscripts zero indicate that this is a neutral process. Each of the four factors in these expressions represents a stage in picking an organism to reproduce and an organism to die. The first term in the sums is the probability of choosing island $j$, denoted by $f_j$, the second is the probability of picking an organism on island $j$ to reproduce, the third is the probability that the offspring of this organism migrates to island $i$ ($m_{ij}$, $i \neq j$) or that it stays on island $j$ ($m_{jj}$), and finally the fourth is the probability of picking the organism to be replaced by the offspring. Note that since the population of each island is assumed fixed, the process of birth/death is necessarily coupled to migration. We now wish to extend this formalism to account for selection and mutation. 

Let us first consider mutation. It is worthwhile noting that there is more than one way that mutation can be modeled~\cite{blythe_mckane_models_2007}; here we will assume that the mutation events are independent of the birth/death events. We allow birth/death/migration events to happen a fraction $b$ of the time, and mutation events a fraction $(1-b)$ of the time. If we denote $\kappa_{1 i}$ to be the mutation rate from $Y$ to $X$ on island $i$ and $\kappa_{2 i}$ to be the mutation rate from $X$ to $Y$ on island $i$, then the transition rates are
\begin{eqnarray}
T_{M}( n_{i}+1 | n_{i} ) &=&  b T_{0}( n_{i}+1 | n_{i} ) + (1-b)\kappa_{1 i}\frac{\beta_i N - n_i}{\beta_i N}\,, \nonumber \\
T_{M}( n_{i}-1 | n_{i} ) &=&  b T_{0}( n_{i}-1 | n_{i} ) + (1-b)\kappa_{2 i}\frac{n_i}{\beta_i N}\,, \label{eq_rates_mutation}
\end{eqnarray}
where the rates $T_{0}( n_{i} \pm 1 | n_{i} )$ are given by \eref{eq_neutral_rates} and where the subscript $M$ denotes `mutation'. We have allowed for the mutation rates to vary from island to island. The concept of mutation varying with habitat is perhaps less intuitive than that of selective pressure changing according to the environment. However, there have been experimental studies of certain species that suggest that mutation rates can increase as a result of external environmental stress factors (see, for example~\cite{mexican_fish}). Note that since there is no selective pressure, the probability of birth/death/migration events is still proportional to the neutral transition rates, albeit moderated by a factor $b$.

We now turn to how selection is incorporated into the metapopulation Moran model. If a well-mixed system features selection, it is assumed that the probability of each of the types reproducing is weighted by some factor, rather than simply being proportional to the frequencies of the respective types. These weightings increase or decrease the propensity of the individuals to reproduce with respect to one another, which generates a selective pressure. Here, these weightings are given by the vectors $\bm{W}_{X}$ and  $\bm{W}_{Y}$ whose $i^{\rm th}$ elements give the fitness weightings on the $i^{\rm th}$ island of types $X$ and $Y$ respectively. The transition rates for the metapopulation model with selection are then
\begin{eqnarray}
 T_{S}( n_{i}+1| n_{i} ) &=& \sum_{j=1}^{\mathcal{D}}  f_{j}  \frac{ [W_{X}]_{j}n_{j}}{[W_{X}]_{j}n_{j}+  [W_{Y}]_{j}(\beta_{j}N-n_{j})}   \times \nonumber \\  & & m_{ij} \frac{(\beta_i N-n_{i})}{\beta_{i}N - \delta_{ij}}   , \nonumber \\
 T_{S}(n_{i}-1|  n_{i} ) &=& \sum_{j=1}^{\mathcal{D}}  f_{j}  \frac{ [W_{Y}]_{j}(\beta_{j}N- n_{j})}{[W_{X}]_{j}n_{j}+  [W_{Y}]_{j}(\beta_{j}N-n_{j})}   \times \nonumber \\ & & m_{ij} \frac{n_{i}}{\beta_{i}N-\delta_{ij}}    \,, \label{eq_rates_selection}
\end{eqnarray}
where the normalization of the fitness terms has been chosen to keep the birth/death rate of the system fixed.

Finally, we wish to include both selection and mutation. Since mutation and selection are independent processes, we would still expect the mutation to be controlled by the term with the prefactor $(1-b)$ in \eref{eq_rates_mutation}. However, we would expect the birth/death/migration events, moderated by the factor $b$, to now include the selective pressures described by the rates \eref{eq_rates_selection}. This leads to the transition rates
\begin{eqnarray*}
T_{MS}( n_{i}+1 | n_{i} ) &=&  b T_{S}( n_{i}+1 | n_{i} ) + (1-b)\kappa_{1 i}\frac{\beta_i N - n_i}{\beta_i N}\,, \nonumber \\
T_{MS}( n_{i}-1 | n_{i} ) &=&  b T_{S}( n_{i}-1 | n_{i} ) + (1-b)\kappa_{2 i}\frac{n_i}{\beta_i N}\,.
\end{eqnarray*}
Since we are free to rescale time in the master equation by a factor of $b$, and absorb a factor of $(1-b)/b$ into the mutation rates, this expression for the transition rates can be simplified to 
\begin{eqnarray}
T_{MS}( n_{i}+1 | n_{i} ) &=& T_{S}( n_{i}+1 | n_{i} ) + \kappa_{1 i}\frac{\beta_i N - n_i}{\beta_i N}\,, \nonumber \\
T_{MS}( n_{i}-1 | n_{i} ) &=& T_{S}( n_{i}-1 | n_{i} ) + \kappa_{2 i}\frac{n_i}{\beta_i N}\,. \label{eq_rates_mutation_selection}
\end{eqnarray}
These transition rates, along with the master equation~(\ref{master}), define a metapopulation Moran model with migration, mutation and selection.

We have now constructed a model containing migration, mutation and selection which, being stochastic, also takes account of genetic drift. However, the master equation is a set of $\mathcal{D}(N + 1)$ difference equations with which it is very difficult to make analytic progress. However, a standard way to simplify the system is available via the diffusion approximation. The assumption at the heart of the diffusion approximation is that, for large enough system size (in this case the island size $\beta_{i}N$), the system can be described by a set of approximately continuous variables $x_{i} = n_{i}/\beta_{i}N$. If additionally the transitions between states are sufficiently local, in this case guaranteed by the fact that transitions from a state $n_{i}$ only take the system to neighboring states $n_{i+1}$ and $n_{i-1}$, a Taylor expansion of the master equation in the continuous variables $x_{i}$ can be be conducted in the small parameter $(\beta_{i}N)^{-1}$.

The full procedure is described mathematically in \Aref{app_diffusion}. Here we simply note that using Eqs.~(\ref{F_defn}) and (\ref{AandB}), together with the transition rates~(\ref{eq_rates_mutation_selection}), the master equation for the system with migration, selection and mutation can be approximated by the Fokker-Planck equation
\begin{eqnarray}
 \frac{\partial p(\bm{x},t)}{\partial t} = &-& \frac{1}{N} \sum_{i=1}^{\mathcal{D}} \frac{\partial}{\partial x_{i}} \left[A_{i}(\bm{x})p(\bm{x},t)\right] \nonumber \\ &+& \frac{1}{2N^{2}}\sum_{i=1}^{\mathcal{D}}\frac{\partial^{2}}{\partial x_{i}^{2}} \left[B_{ii}(\bm{x})p(\bm{x},t)\right].
\label{FPE}
\end{eqnarray}
The drift vector has elements
\begin{eqnarray}
& & A_{i}(\bm{x}) = \frac{1}{\beta_{i}}\sum_{j = 1 }^{\mathcal{D}} \frac{ G_{ij} }{ ([W_{X}]_{j} - [W_{Y}]_{j})x_{j} + [W_{Y}]_{j} } \times 
\nonumber \\ 
& & \left\{ [W_{X}]_{j} x_{j} - [W_{Y}]_{j} x_{i} - ( [W_{X}]_{j} - [W_{Y}]_{j} ) x_{i} x_{j} \right\} \nonumber \\
& & + \frac{1}{\beta_{i}} \left[ \kappa_{1i} - ( \kappa_{1i} + \kappa_{2i} ) x_{i} \right] ,\, \label{A_general}
\end{eqnarray}
%where by $\mathcal{O}(\kappa^2)$ we mean any product of two or more of the $\kappa_{1i}$ or $\kappa_{2i}$, $i=1,\ldots,\mathcal{D}$, 
and the diffusion matrix is
\begin{eqnarray}
 B_{ii}(\bm{x}) = \frac{1}{\beta_{i}^{2}}\sum_{j = 1 }^{\mathcal{D}} \frac{ G_{ij} }{ ([W_{X}]_{j} - [W_{Y}]_{j}) x_{j} + [W_{Y}]_{j} } \times \nonumber \\ 
\left\{ [W_{X}]_{j} x_{j} + [W_{Y}]_{j} x_{i} - ( [W_{X}]_{j} + [W_{Y}]_{j} ) x_{i} x_{j} \right\} + \mathcal{O}( \bm{\kappa}_{1} , \bm{\kappa}_{2} ) \nonumber \\ \label{B_general}
\end{eqnarray}
where the vectors $\bm{\kappa}_{1}=(\kappa_{11},\kappa_{12},\ldots,\kappa_{1\mathcal{D}})$ and $\bm{\kappa}_{2}=(\kappa_{21},\kappa_{22},\ldots,\kappa_{2\mathcal{D}})$ have been introduced and where $G_{ij} \equiv m_{ij}f_{j}$. The parameters 
$\kappa_{1i}$ and $\kappa_{2i}$ are assumed to be small, of the order of $N^{-1}$, so that the order $\bm{\kappa}_{1}$ and $\bm{\kappa}_{2}$ terms in \eref{B_general} are of the same order as the $N^{-3}$ terms neglected in the expansion of the master equation, and so may be similarly neglected.

In what follows, we will make use of the equivalence~\cite{gardiner_2009,risken_1989} between the FPE (\ref{FPE}) and the It\={o} stochastic differential equation (SDE) 
\begin{equation}\label{generalSDE}
\dot{x}_{i} = A_{i}(\bm{x}) + \frac{1}{\sqrt{N}}\eta_{i}(\tau) \, ,
\end{equation}
where the dot indicates differentiation with respect to $\tau=t/N$, and $\bm{\eta}(\tau)$ is a Gaussian white noise with zero mean and correlation functions
\begin{equation}
\langle \eta_{i}(\tau)\eta_{j}(\tau') \rangle = B_{ij}(\bm{x}) \delta(\tau-\tau') \, . 
\label{eta_correlator}
\end{equation}
It is useful to think of this intuitively as the deterministic system, $\dot{x}_{i} = A_{i}(\bm{x})$, with a small amount of added noise.  With the full details of the Fokker-Planck equation now in hand, we may proceed to analyze the behavior of the system.

\section{Removing fast degrees of freedom from the metapopulation model with mutation}\label{sec_mutation}

To begin our analysis, we consider first the metapopulation Moran model with mutation but no selection; $[W_{X}]_{i}=[W_{Y}]_{i}=0$ for each island. Once again the FPE for this system is given by \eref{FPE} but now with the elements of the drift vector taking the simplified form
\begin{eqnarray}
 A_{i}(\bm{x}) = \sum^{\mathcal{D}}_{j=1} H_{ij} x_{j} + \frac{1}{\beta_{i}} \left[ \kappa_{1i} - ( \kappa_{1i} + \kappa_{2i} ) x_{i} \right] \, , 
\label{A_mutation}
\end{eqnarray}
where the elements of the matrix $H$ are given by \eref{defineH}. The diagonal elements of the diffusion matrix meanwhile are given by 
\begin{eqnarray}
B_{ii}(\bm{x}) &=& \frac{1}{\beta_{i}^{2}}\sum_{j=1}^{\mathcal{D}}G_{ij}\left(  x_{i} + x_{j} - 2x_{i}x_{j}\right)  \,,
\label{B_mutation}
\end{eqnarray}
where the higher order terms $\bm{\kappa}_{1}$ and $\bm{\kappa}_{2}$ in \eref{B_general} have been neglected.

\begin{figure}
\includegraphics[width=0.4\textwidth]{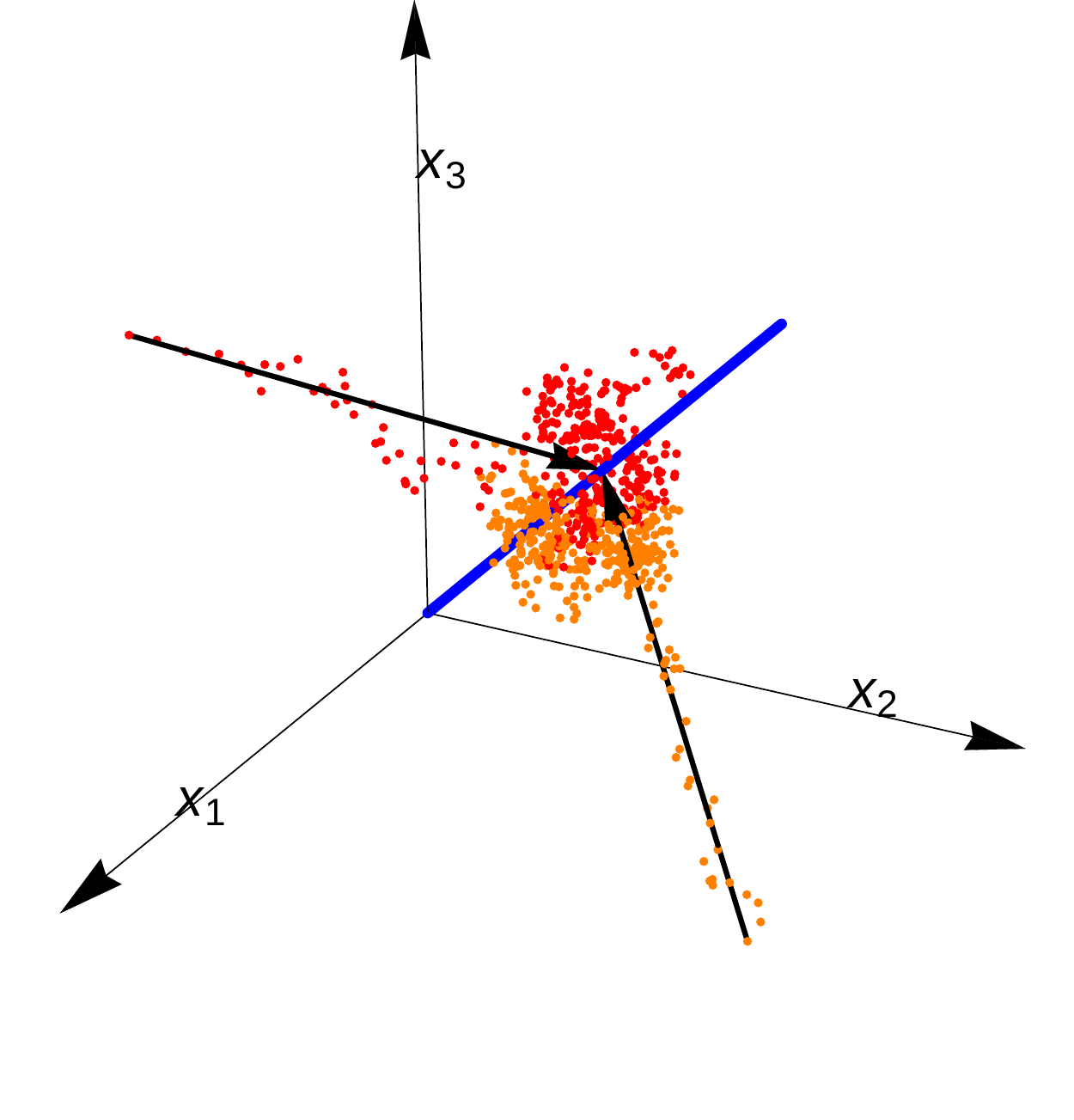}
\caption{(Color online)  Trajectories of a neutral system with three islands. It can be seen that after a short time, the stochastic trajectories denoted by orange and red points, come to lie in the region of the one-dimensional center manifold, the thick blue line $x_{1}=x_{2}=x_{3}$. The stochastic trajectories follow an approximately deterministic trajectory to the center manifold, indicated by the black arrows.}
\label{fig_3D_traj}.
\end{figure} 

To understand how fast-variable elimination can be used to simplify this system, it is best to briefly consider the neutral deterministic system obtained by taking $\kappa_{1i}=\kappa_{2i}=0$ for all $i$, and $N\rightarrow\infty$. In this case, the deterministic system is linear: $\dot{x}_{i}=\sum_j H_{ij}x_j$, and has a very simple dynamics. This is a consequence of the fact that the eigenvalue spectrum of the matrix $H$ is such that its largest eigenvalue, $\lambda^{(1)}$ is zero, and the remaining $\mathcal{D}-1$ eigenvalues, $\lambda^{(i)}$, $i=2,\ldots,\mathcal{D}$, have a negative real part~\cite{constable_phys}. Therefore, as long as these negative real parts are not too close to zero, the dynamics will consist of a rapid collapse of $\mathcal{D}-1$ `fast' variables, onto the center manifold, defined by the right-eigenvector of $H$ corresponding to the zero eigenvalue. The equation for this line is given by $x_{i} = x_{j}$ for all $i$ and $j$ (see \Aref{app_remove_fast}). The remaining eigenvectors are used to determine the fast directions. For finite $N$ there is still a collapse to the center manifold which is dominated by the deterministic dynamics, but near to the center manifold, the dynamics consists of stochastic drift along the center manifold, until an axis is reached, and fixation occurs. An illustrative case for three islands is presented in \fref{fig_3D_traj}.

\begin{figure}
\includegraphics[width=0.4\textwidth]{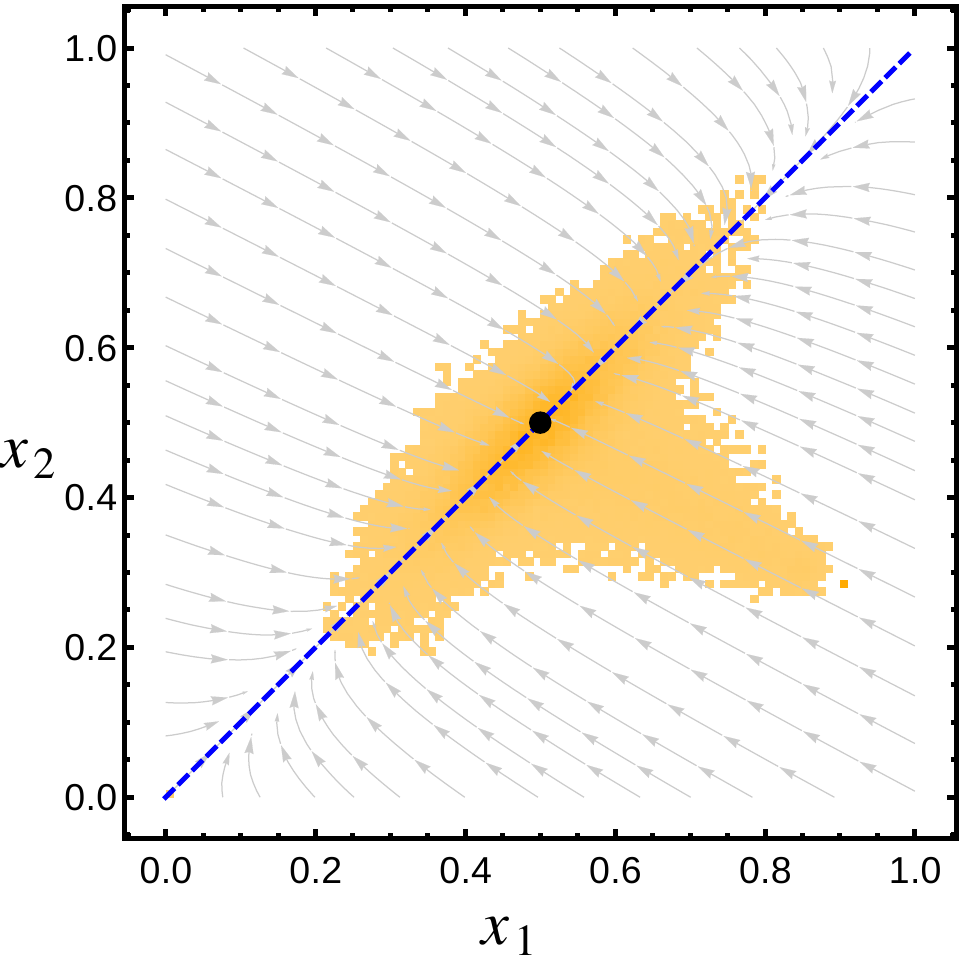}
\caption{(Color online) Deterministic trajectories (gray) plotted for the two-island Moran model with mutation. The slow subspace, \eref{general_slow_subspace}, is plotted as a blue dashed line. A histogram of trajectories of the original IBM are overlaid in orange. Having relaxed to the slow subspace, they can be seen to be confined to its vicinity.}
\label{fig_traj_m}.
\end{figure}

Since $\kappa_{1i}$ and $\kappa_{2i}$ are assumed to be small, we still expect the separation between the fast and slow timescales to hold when mutation is included, which should allow us to eliminate the $\mathcal{D}-1$ fast modes, leaving an effective model with one degree of freedom. That this is the case is illustrated in \fref{fig_traj_m}. However now there is no center manifold, but instead a slow subspace. This leaves us with a problem; without a line of fixed points about which we can linearize, there are no eigenvectors with which to characterize the fast and slow modes. Our solution to this difficulty, which worked extremely well when selection was added~\cite{constable_phys}, is to continue to use the eigenvectors found when mutation is absent. Since the mutation rates are very small, this should, and as we will see does, provide a reduced model which is an excellent approximation to the original IBM.

To recap then, the fast-mode elimination procedure consists of eliminating the dynamics in the $\mathcal{D}-1$ fast directions, and keeping only the dynamics in the `slow' direction. There is more than one way to carry this out, as will be discussed in \sref{sec_relate}. While a more detailed summary of the method discussed in \cite{constable_phys} and \cite{constable_bio} is given in \Aref{app_remove_fast}, here we will content ourselves with stating the main features of the procedure.

The method makes extensive use of the left- and right-eigenvectors of $H$, denoted $\bm{u}^{(i)}$ and $\bm{v}^{(i)}$ respectively. In the case of the neutral system, $\bm{v}^{(1)}$ is coincident with the slow (stationary) direction, while the vectors $\bm{v}^{(j)}$ for $j \geq 2$ give information about the fast directions. The vectors $\bm{u}^{(i)}$ meanwhile form an orthogonal set such that
\begin{equation}
\sum^{\mathcal{D}}_{k=1}\,u^{(i)}_k v^{(j)}_k = \delta_{i j}.
\label{orthonormal}
\end{equation}
The method we use in this paper consists of two separate steps. The first is that the system is restricted to the slow subspace of the deterministic system. Obtaining an analytic description for this line is not straightforward, however a very good approximation can be obtained as the solution to the equation
\begin{equation}\label{general_slow_subspace}
\sum_{i=1}^{\mathcal{D}} u^{(j)}_{i} A_{i}(\bm{x}) = 0 \, , 
\qquad j = 2, \ldots, \mathcal{D}.
\end{equation}
Here the slow subspace is approximated as the space on which the drift vector has no components in the fast directions of the neutral system. To complete the reduction procedure, we apply a projection matrix to the SDE (evaluated on the slow subspace), to remove any further contributions from the fast directions. The projection matrix in this case takes the particularly simple form
\begin{eqnarray}\label{defineProjection}
 P_{ij} =  u^{(1)}_{j}\, , \qquad  \, i = 1, \ldots , \mathcal{D}.
\end{eqnarray}
Essentially this maps the deterministic dynamics on the slow subspace onto the line $x_{i} = z$ while simultaneously removing any contribution from the noise in the fast directions.

We now wish to apply this method to the SDE~(\ref{generalSDE}) with drift vector \eref{A_mutation} and diffusion matrix \eref{B_mutation}. We begin by decomposing the coordinate $x_i$ as follows:
\begin{equation}
x_i = z v^{(1)}_i + \sum^{\mathcal{D}}_{j=2} w_{j} v^{(j)}_{i},\ i=1,\ldots,\mathcal{D},
\label{decompose}
\end{equation}
where $z$ is the slow mode, the $w_j$ are the fast modes. If we now apply the condition in \eref{general_slow_subspace}, but to $A_{i}(\bm{x})$ given by \eref{A_mutation}, we find in terms of the coordinates $z$ and $w_j$ that
\begin{equation}
w_j = - \frac{1}{\lambda^{(j)}} \sum^{\mathcal{D}}_{k=1} \frac{u^{(j)}_{k}}{\beta_k} \left[ \kappa_{1k} - (\kappa_{1k} + \kappa_{2k}) z \right] + 
\mathcal{O}(\kappa^2),
\label{slow_variables_mut}
\end{equation}
where by $\mathcal{O}(\kappa^2)$ we mean any product of two or more of the $\kappa_{1i}$ or $\kappa_{2 i}$, $i=1,\ldots,\mathcal{D}$. In essence, Eqs.~(\ref{decompose}) and (\ref{slow_variables_mut}) tell us the following. When there is no mutation present, the deterministic system rapidly collapses to the center manifold $x_i = z$, for all $i$. When mutation is introduced, the collapse still happens, but now it is to the slow subspace, which from Eqs.~(\ref{decompose}) and (\ref{slow_variables_mut}) is seen to be given by
\begin{equation}
x_i = z - \sum^{\mathcal{D}}_{j=2} \sum^{\mathcal{D}}_{k=1} \frac{u^{(j)}_k v^{(j)}_{i}}{\lambda^{(j)} \beta_k} \left[ \kappa_{1k} - (\kappa_{1k} + \kappa_{2k}) z \right] ,
\label{slow_subspace_mut}
\end{equation}
where $i=1,\ldots,\mathcal{D}$ and where we have again neglected $\mathcal{O}(\kappa^2)$ terms. Since in practice the mutation rates are tiny, the deviation of the slow subspace defined in \eref{slow_subspace_mut} from the the line $x_i = z$ is almost impossible to see on any plot. Furthermore, to calculate quantities to first order in the mutation strength the specific form of \eref{slow_subspace_mut} is not required, as we shall see below. However, we will use the concept of a slow subspace in a more substantive way in \sref{sec_mutation_selection}, when selection is introduced.

Therefore the application of the fast mode elimination procedure now first consists of evaluating $A_{i}(\bm{x})$ and $B_{ii}(\bm{x})$ on the slow subspace, and then using the projection operator (\ref{defineProjection}), to obtain $\bar{A}(z) = \sum_{i=1}^{\mathcal{D}} u^{(1)}_{i} A_{i}(z)$ and $\bar{B}(z)=\sum_{i=1}^{\mathcal{D}} \left[ u^{(1)}_{i}\right]^{2} B_{ii}(z)$, the effective drift and diffusion terms on the center manifold. The first term in Eq.~(\ref{A_mutation}) does not contribute, since when acted upon by $u^{(1)}_i$ it vanishes. In the second term we only need to set $x_i$ equal to $z$ to this order, and so using the definition of $\bar{A}(z)$ we find 
\begin{equation}
\bar{A}(z) = \sum^{\mathcal{D}}_{i=1} \frac{u^{(1)}_{i}}{\beta_i} 
\left[ \kappa_{1i} - (\kappa_{1i} + \kappa_{2i}) z \right] + 
\mathcal{O}(\kappa^2)\,.
\label{Abar_mut}
\end{equation}
To this order $\bar{B}(z)$ is identical to that found without mutation (see Eqs.~(\ref{defineBbar_app})-(\ref{b_1})), that is $\bar{B}(z) = 2 b_{1} z (1 - z)$
with
\begin{equation}
b_{1} = \sum_{i,k=1}^{\mathcal{D}}\,[u^{(1)}_{i}]^{2}G_{ik}\beta_{i}^{-2}.
\label{define_b_1}
\end{equation}

Therefore the reduced FPE is
\begin{equation}
\frac{\partial p}{\partial \tau} = - \frac{\partial }{\partial z} \left[ \bar{A}(z) p \right] + \frac{1}{2N} \frac{\partial^2 }{\partial z^2} \left[ \bar{B}(z) p \right],
\label{reduced_FPE}
\end{equation}
where
\begin{equation}
\bar{A}(z) = \hat{\kappa}_1 - \left( \hat{\kappa}_1 + \hat{\kappa}_2 \right)z,
\ \ 
\bar{B}(z) = 2b_1 z\left( 1 - z \right),
\label{Abar_Bbar_mut}
\end{equation}
to leading order in $\kappa_{1i}$ and $\kappa_{2i}$. Here $b_1$ is given by \eref{define_b_1} and 
\begin{equation}
\hat{\kappa}_1 = \sum^{\mathcal{D}}_{i=1} \frac{u^{(1)}_i \kappa_{1i}}{\beta_i}, \ \
\hat{\kappa}_2 = \sum^{\mathcal{D}}_{i=1} \frac{u^{(1)}_i \kappa_{2i}}{\beta_i}.
\label{kappa_hats_defn}
\end{equation}

The drift and diffusion coefficients of the effective model, given by Eq.~(\ref{Abar_Bbar_mut}) have precisely the form of a well-mixed Moran model, but with a noise strength enhanced by a factor of $b_1$ and mutation rates expressed by Eq.~(\ref{kappa_hats_defn}). Even if the mutation rates do not vary from island to island, they will still be enhanced by a non-trivial factor of $\sum_i u^{(1)}_i/\beta_i$. 

As we have already stressed, having non-zero mutation rates ensures that the system never reaches fixation. Instead it will eventually approach a stationary pdf which, because the FPE is that of a one-island model, has the well-known form~\cite{roughgarden_1979,crow_1970,ewens_2004}
\begin{equation}
p_{\rm st}(z) = \mathcal{N}z^{c_1} \left( 1 - z \right)^{c_2},
\label{spdf_mut}
\end{equation}
where $\mathcal{N}$ is a normalization constant and where
\begin{equation}
c_1 = \frac{N}{b_1}\hat{\kappa}_1 - 1, \ \ 
c_2 = \frac{N}{b_1}\hat{\kappa}_2 - 1.
\label{c_mut}
\end{equation}
The effect of the structure of the network of islands on the dynamics can now be investigated.

\begin{figure}[t]
\includegraphics[width=0.44\textwidth]{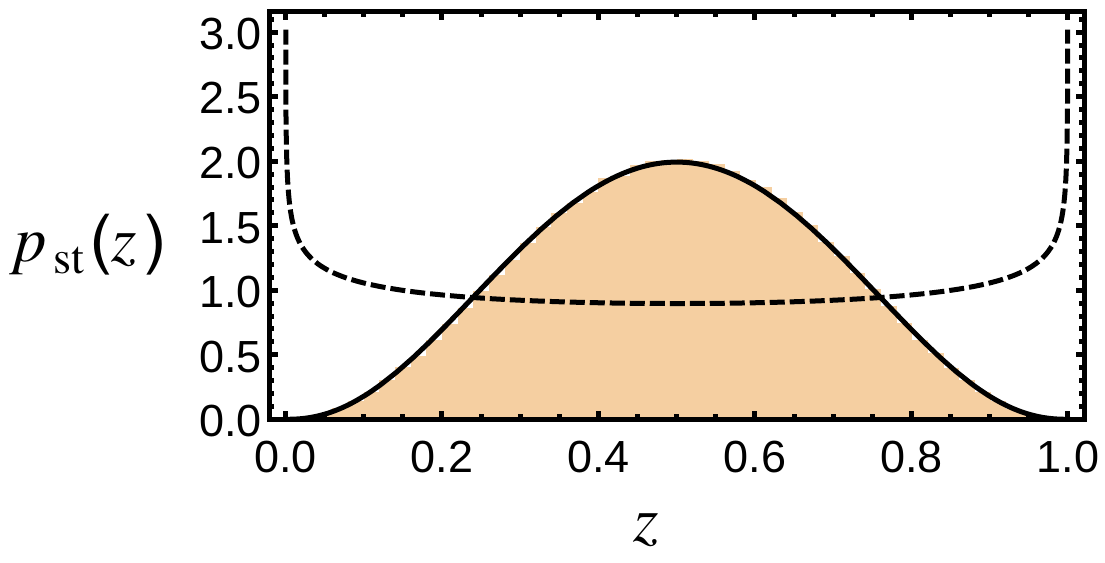}
\caption{(Color online) Stationary pdf of a metapopulation Moran model with mutation. The system has four islands of equal size $N=300$ connected by a symmetric migration matrix with diagonal entries $m_{ii}=0.9$. The mutation rates are $\kappa_{1i}=\kappa_{2i}=7\times10^{-4}$ for each island. The filled orange histogram is obtained from stochastic simulations of the IBM, while the solid black line is obtained from theory (\eref{spdf_mut}). The dashed line is the theoretical prediction of a well-mixed model with the same average mutation rates and a population size $\mathcal{D}N$.}
\label{fig_destroy_bistable}
\end{figure} 

In order to examine the effect of the metapopulation structure, the predictions of the effective models will be compared with those obtained from a well-mixed, unstructured analog. The unstructured analog is taken to be a well-mixed one-island model with the mean mutation rates of the metapopulation model (weighted by island size). We begin by considering the most simple case, that when all islands are the same size, $\beta_{i}=1$ $\forall\, i$, and where the matrix $G$ is symmetric. If $G$ is  symmetric it is straightforward to show~\cite{constable_phys} that $u_i^{(1)}$ is given by $\beta_i/\sum_j \beta_j$ and $b_{1}$ by $(\sum_j \beta_j )^{-2}$. Therefore if in addition $\beta_i = 1$, one obtains $\hat{\kappa}_{1} = \sum_{j=1}^{\mathcal{D}} \kappa_{1j}/\mathcal{D}$ and $\hat{\kappa}_{2} = \sum_{j=1}^{\mathcal{D}} \kappa_{2j}/\mathcal{D}$. The effective mutation rates are simply equal to the mean of the mutation rates across demes. The effect of the term $N/b_{1}$ in Eq.~(\ref{c_mut}) is not so straightforward. Recalling that the total population is given by $N_{\rm{Tot}}=N\mathcal{D}$ in this situation, one finds $N/b_{1} = \mathcal{D}N_{\rm{Tot}}$. The reduced system therefore has a greater effective system size than its well-mixed unstructured analog. This means that, in the case where the matrix $G$ is symmetric, the effect of the population structure is to reduce the effect of the noise on the stationary distribution. An example of such a case is given in Fig.~\ref{fig_destroy_bistable}. The noise induced bistability found in the well-mixed unstructured analogue of the model~\cite{maruyama_1977} is no longer present, and the deterministic dynamics dominate.

\begin{figure}[t]
\includegraphics[width=0.44\textwidth]{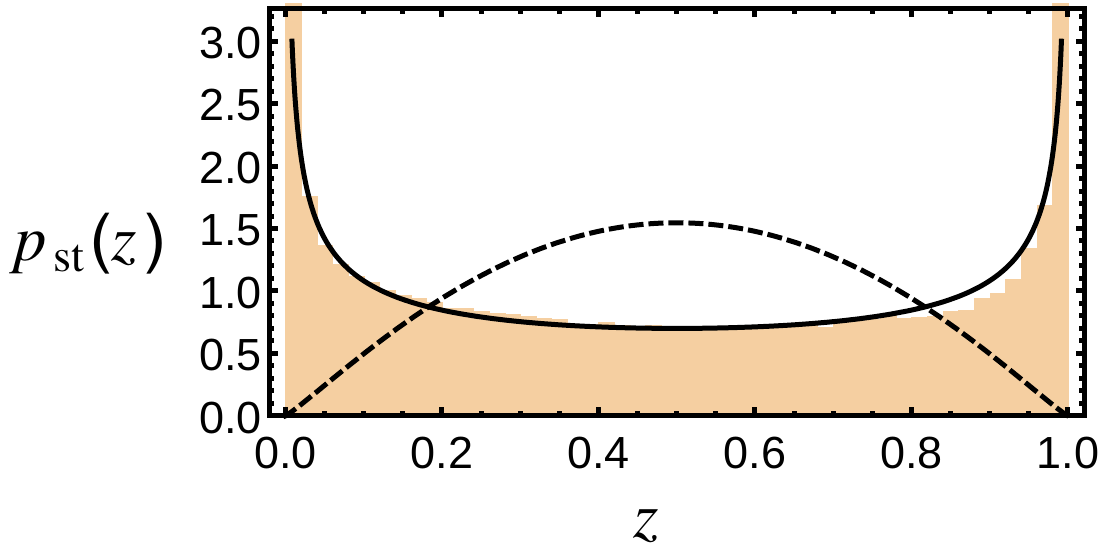}
\caption{(Color online) Similar plots to \fref{fig_destroy_bistable} but for a two-island system with asymmetric migration, differing islands sizes and varying mutation rates. While for brevity most parameters are not stated here, we note that the mutation rates differ by many magnitudes over the islands (see \Aref{app_parameters}). %$\bm{\kappa}_{1}=\kappa_{2}=( 1\times10^{-5}, 5\times10^{-3} )$. 
Numerical investigations suggest this is necessary in order to enhance the bistability of the system relative to the well-mixed unstructured analog (with the same total population size and mean mutation rates) the pdf of which is here plotted as the dashed black line. }
\label{fig_enhance_bistable}
\end{figure}

\begin{figure}[t]
\includegraphics[width=0.48\textwidth]{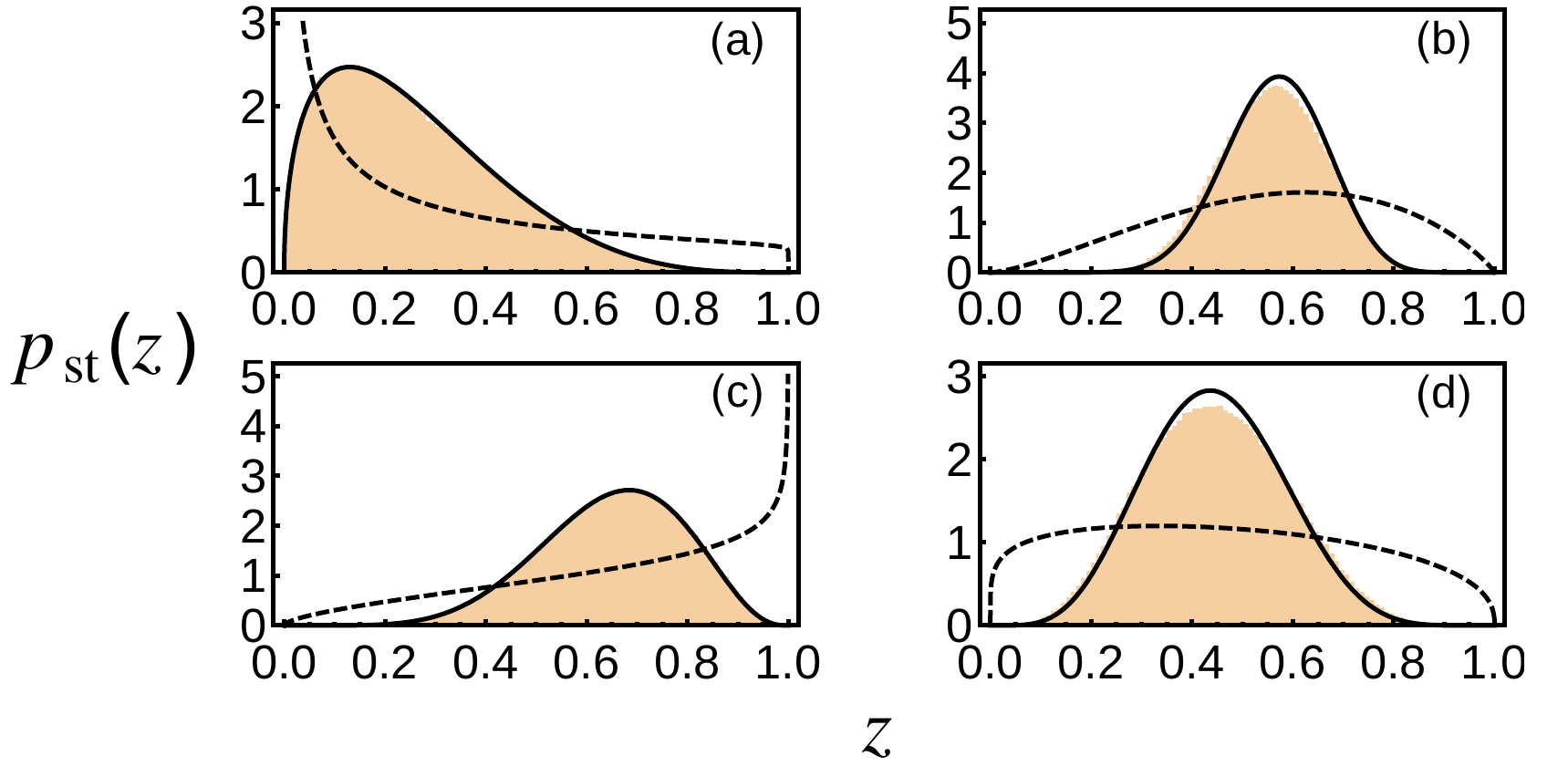}
\caption{(Color online) The stationary pdf on the slow subspace for a range of systems with various parameters which are omitted here for brevity but which can be found in \Aref{app_parameters}. Once again the solid black line is obtained from theory, the orange histogram from simulations of the original IBM, and the dashed line from a well mixed model with the same total system size and average mutation rates (weighted by island size).}
\label{fig_range_parameters}
\end{figure}

In general, it is found that the effect of population structure identified above is seen in most other parameter regimes. That is, the effect of population structure is in general to reduce the effect of the noise on the stationary distribution, relative to a well-mixed system with the same total population size and mean mutation rates. However, there do exist some cases where the converse is true, where the population structure \emph{increases} the effect of noise relative to the well-mixed model. In Fig.~\ref{fig_enhance_bistable}, the stationary distributions for a well-mixed model and a two-island system are plotted for a situation in which this is the case. We note that numerically it appears that such behavior is only possible if the elements of $\bm{\kappa}_{1}$ and $\bm{\kappa}_{2}$ are allowed to vary significantly across demes. More investigation is clearly needed to explore the full range of behavior possible in this system. However, as demonstrated in Fig.~\ref{fig_range_parameters}, the analytic predictions derived from the reduced model provide remarkably good agreement with the results from Gillespie simulation of the full model, across a range of parameters.

\section{The reduced model with mutation and selection}\label{sec_mutation_selection}

We now proceed to obtain a reduced-dimension description for the model with mutation and selection. We have applied fast mode elimination techniques to Moran metapopulation models with selection elsewhere~\cite{constable_phys,constable_bio}, and we will use many of the results previously obtained in the analysis we carry out here. There are, however, some ways in which our previous analysis and the current one differ. For instance, here we will include mutation as well as selection. Since we treat these as independent processes, they can be added separately, at least to the order at which we are working. However, as in Sec.~\ref{sec_mutation}, the alleles will not fix, and the focus will instead be on the stationary pdf of the FPE (\ref{reduced_FPE}). 

We recall that the model with mutation and selection is approximated by the FPE~(\ref{FPE}) with drift and diffusion elements specified by Eqs.~(\ref{A_general}) and (\ref{B_general}) respectively. Note that we have not as yet made any assumption about the size of the selection weightings $[W_{X}]_{i}$ and $[W_{Y}]_{i}$; the diffusion approximation can be made for arbitrary selection strengths. However, as seen in the previous section, the fast-variable elimination technique which we seek to implement relies on small departures from the neutral (linear) model. To make progress, we therefore wish to consider the limit of weak selection. In this spirit, we assume that the relative fitness of alleles $X$ on island $i$ is $[W_{X}]_i = 1 + s\rho_i$, and correspondingly $[W_{Y}]_i = 1 + s\sigma_i$ for allele $Y$ on island $i$, where the strength of selection parameter, $s$, is small. This particular choice for the weightings gives another difference from our earlier work, in which we took $\sigma_i = 0$ (and denoted $\rho_i$ as $\alpha_i$)~\cite{constable_phys,constable_bio}. We will find that the choice made in this paper is a less restrictive assumption concerning the nature of the relative finesses of alleles $X$ and $Y$.

Since the parameter $s$ is assumed to be small, having substituted the fitness weightings $[W_{X}]_i = 1 + s\rho_i$, and $[W_{Y}]_i = 1 + s\sigma_i$ into \eref{A_general} and \eref{B_general}, a Taylor expansion of the resulting expressions can be conducted in $s$. One then obtains
\begin{eqnarray}
& & A_{i}(\bm{x}) = \sum_{j=1}^{\mathcal{D}}H_{ij}x_{j} +  \frac{1}{\beta_{i}}\left[ \kappa_{1i} - (\kappa_{1i} + \kappa_{2i}) x_{i} \right] \nonumber \\
& & + \frac{s}{\beta_i} \sum_{j=1}^{\mathcal{D}} G_{ij} x_j \left( 1 - x_j \right) \left( \rho_j - \sigma_j \right) \left[ 1 - s\sigma_j - s\left( \rho_j - \sigma_j \right) x_j \right]\,,
\nonumber \\
\label{A_general_s}
\end{eqnarray}
to the order we are working, and with 
\begin{equation}
B_{ii}(\bm{x}) = \frac{1}{\beta_{i}^{2}}\sum_{j=1}^{\mathcal{D}}G_{ij}\left(  x_{i} + x_{j} - 2x_{i}x_{j}\right) \,,
\label{B_ii}
\end{equation}
which takes the same form as the neutral model, again to the order we are working. The phrase `to the order we are working' in the case of the drift coefficient means that we neglect order $s^3$ terms. This is because we either assume that $s$ is of order $N^{-1}$ or smaller, or alternatively of the order of $N^{-1/2}$ or smaller. In addition, we will always assume that $\kappa_{1i}$ and $\kappa_{2i}$ are of the order of $N^{-1}$ or smaller. The motivation for these choices are the small values of these parameters typically found in practice~\cite{lynch2010,nielson2003}. The scaling with $N$ is purely a mathematical convenience however, and while there do exist cases in which the selection strengths of competing genotypes are large~\cite{barret2006}, we do not consider such regimes here. By a similar argument we neglect all corrections to the diffusion matrix which involve powers of $s, \kappa_{1i}$ or $\kappa_{2i}$. 

\begin{figure}[t]
\includegraphics[width=0.48\textwidth]{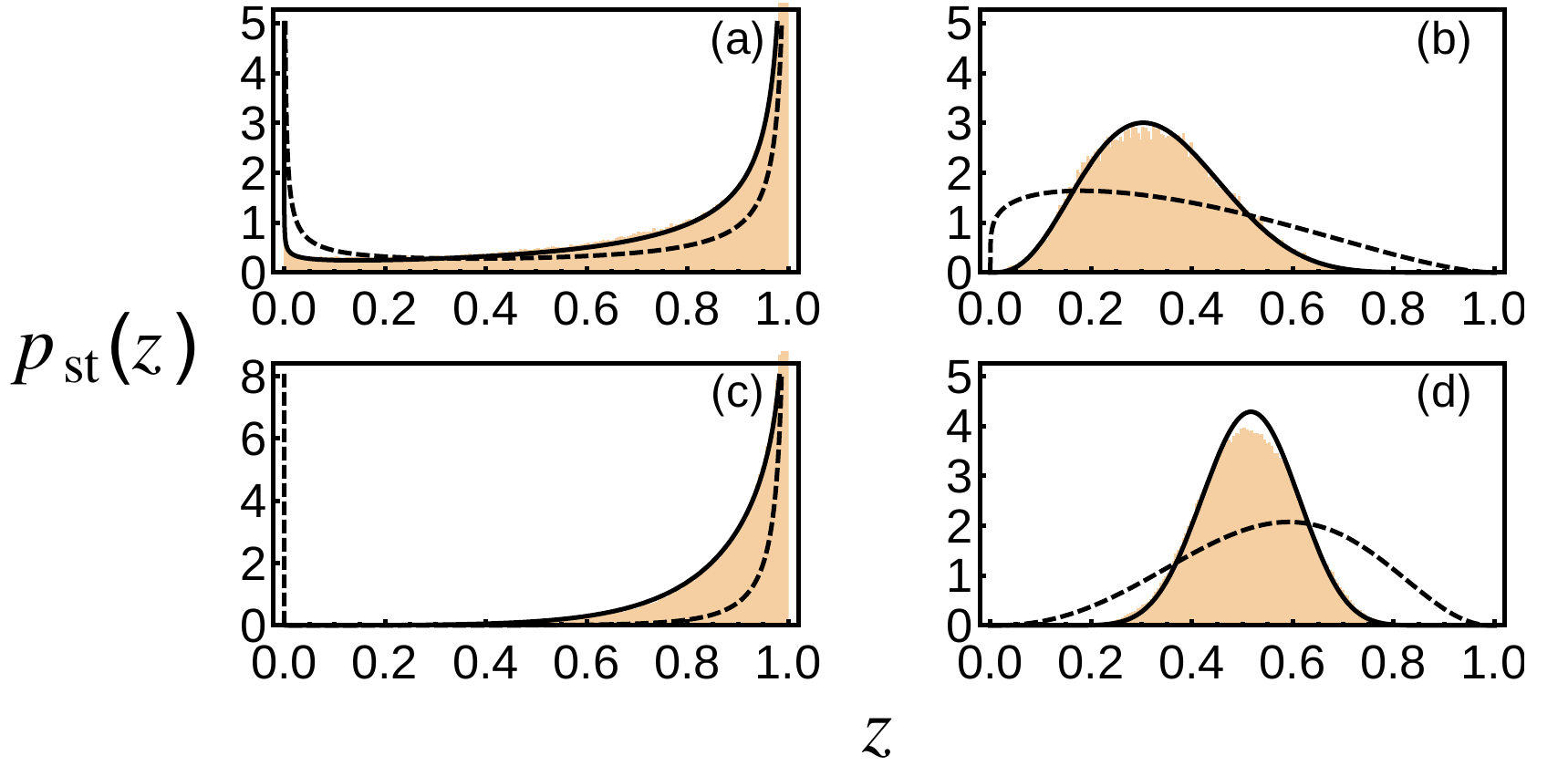}
\caption{(Color online) The stationary pdf in the slow variable $z$ for systems in which both selection and mutation are present. The precise parameters can once again be found in \Aref{app_parameters}. Once more the the orange histogram is obtained from simulations of the original IBM, while the solid black line is obtained from reduced theory. The theory once again matches simulations extremely well, especially in comparison with the predictions of a well-mixed model with the same total system size and average mutation rates weighted by island size (black dashed line).}
\label{fig_selection_range_parameters}
\end{figure} 

We can now apply the fast mode elimination procedure, following the method in \sref{sec_mutation} for mutation without selection. There are some differences however. One is that we are keeping terms of order $s^2$, and so we will need the explicit form for the equation of the slow subspace (the analog of \eref{slow_subspace_mut}, but for selection) to determine $\bar{A}(z)$ to order $s^2$. Another difference from mutation is that the selection terms in \eref{A_general_s} are nonlinear. Nonetheless, as we have already stressed, mutation and selection may be treated as independent processes to the order at which we are working, and the calculation of the equation of the slow subspace and of the resulting form of $\bar{A}(z)$ has already been carried out~\cite{constable_phys}. Taking over these results we find that the fast coordinates $w_j$ may be expressed in terms of the slow coordinate $z$ as
\begin{equation}
w_{j}(z) = - \frac{s z (1 - z) }{\lambda^{(j)}} \sum_{i,k=1}^{\mathcal{D}}\frac{u^{(j)}_{i}G_{ik}\left( \rho_k - \sigma_k \right)}{\beta_{i}} + \mathcal{O}(s^{2},\kappa) \,,
\label{slow_variables_sel}
\end{equation}
where $j=2,\ldots,\mathcal{D}$ and where by $\mathcal{O}(\kappa)$ we mean of order $\kappa_{1i}$ or $\kappa_{2i}$. In fact these order $\kappa$ terms are exactly given by the expression in \eref{slow_variables_sel}, which we have seen in \sref{sec_mutation} is not required to find $\bar{A}(z)$ to first order in $\kappa$.  

Evaluating the drift on the slow subspace, and using the projection operator 
$u^{(1)}_{i}$, as in \sref{sec_mutation} allows us to find $\bar{A}(z)$. The result is the sum of that found in \sref{sec_model} for mutation and that found in Ref.~\cite{constable_phys} for selection, and is given by
\begin{eqnarray}
\bar{A}(z) &=& s a_{1}  z (1 - z) + s^2 \left[ a_3 + a_4 \right] z(1-z) 
\nonumber \\
&+& s^{2} \left[ a_{2} - 2a_3 \right] z^{2}(1-z) 
+ \hat{\kappa}_1 - \left( \hat{\kappa}_1 + \hat{\kappa}_2 \right)z \, , \nonumber \\
\label{Abar-mut_sel}
\end{eqnarray}
with $\bar{B}(z)$ still taking on the neutral form $2b_1 z( 1 - z )$. The constants $a_1, a_2, a_3$ and $a_4$ are given by
\begin{eqnarray}
& & a_{1} = \sum_{i,j=1}^{\mathcal{D}}u^{(1)}_{i} \frac{G_{ij}\alpha_j}{\beta_{i}}\,,
\ \ a_{2} = - \sum_{i,j=1}^{\mathcal{D}}u^{(1)}_{i} \frac{G_{ij}\alpha_j^2}
{\beta_{i}}\,, \nonumber \\
& & a_{3} = - \sum_{m=2}^{\mathcal{D}} \left[\sum_{i,j=1}^{\mathcal{D}}
\frac{u^{(1)}_{i}G_{ij}\alpha_j}{\beta_{i}} \sum_{k,l=1}^{\mathcal{D}} 
\frac{v^{(m)}_{j} u^{(m)}_{k}}{\lambda^{(m)}} 
\frac{G_{kl}\alpha_l}{\beta_{k}} \right]\,, \nonumber \\
& & a_{4} = - \sum_{i,j=1}^{\mathcal{D}}u^{(1)}_{i} \frac{G_{ij}\sigma_j \alpha_j}
{\beta_{i}}\,,
\label{a's}
\end{eqnarray}
where $\alpha_j \equiv \rho_j - \sigma_j$ and with $\hat{\kappa}_1$ and $\hat{\kappa}_2$ being given by Eq.~(\ref{kappa_hats_defn}). Note that while to first order it is only the difference between $\bm{\rho}$ and $\bm{\sigma}$, $\bm{\alpha}$, which impacts the effective strength of selection, at second order the specific forms of $\bm{\rho}$ and $\bm{\sigma}$ are relevant. Though in many applications these second order effects are too small to be noticed, the metapopulation structure allows for migration selection balance, in which the sum of the elements of $\bm{\alpha}$ tend to zero. In cases such as these the precise form of $\bm{\rho}$ and $\bm{\sigma}$ must be determined.

\begin{figure}[t]
\includegraphics[width=0.44\textwidth]{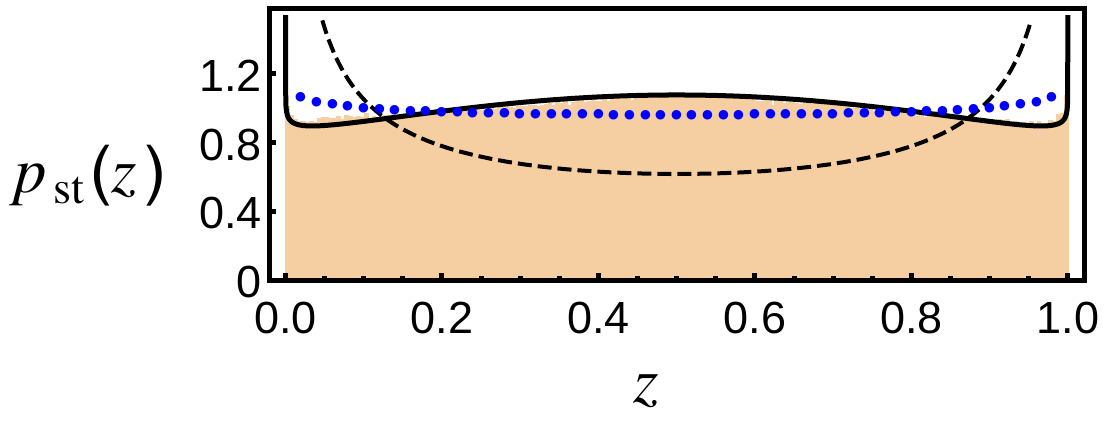}
\includegraphics[width=0.44\textwidth]{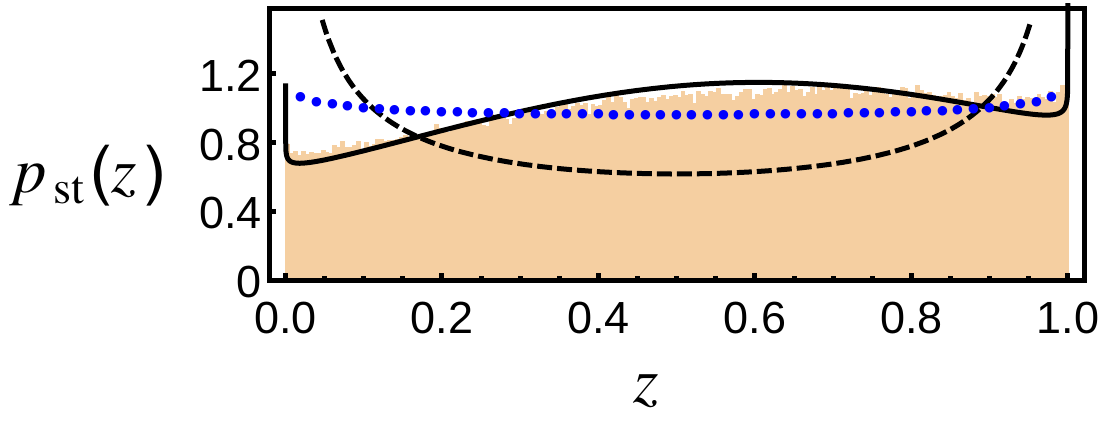}
\caption{(Color online)  Stationary pdfs for two-island systems exhibiting migration-selection balance. In both cases mutation is relatively small (see \Aref{app_parameters}). In these plots, along with the results from simulations (orange histograms), the predictions from the approximation \eref{spdf_mut_sel} (solid black lines) and the predictions from a well-mixed system without the metapopulation structure (black dashed lines), the prediction of \eref{spdf_mut_sel} at first order in $s$ has also been given (blue dotted line). }
\label{fig_mig_selec_balance}
\end{figure}

As in \sref{sec_mutation}, the long-time behavior of the system is encapsulated in the stationary pdf. Mathematically this is found by solving the reduced FPE (\ref{reduced_FPE}) with reflecting boundary conditions, that is, zero probability current at the boundaries. This gives~\cite{gardiner_2009,risken_1989}
\begin{equation}
p_{\rm st}(z) = \frac{1}{\bar{B}(z)}\exp{\int^{z} \frac{2N \bar{A}(z)}{\bar{B}(z)}\,dz}.
\label{gen_stat_pdf}
\end{equation}
Substituting in the explicit form for $\bar{A}(z)$ from Eq.~(\ref{Abar-mut_sel}) and $\bar{B}(z)=2b_1 z(1-z)$, one finds the stationary pdf to be
\begin{equation}
p_{\rm st}(z) = \mathcal{N}z^{c_1} \left( 1 - z \right)^{c_2}\exp{\left( c_3 z
+ c_4 z^2 \right) },
\label{spdf_mut_sel}
\end{equation}
where $\mathcal{N}$ is a normalization constant, $c_1$ and $c_2$ are given by Eq.(\ref{c_mut}) and 
\begin{equation}
c_3 = \frac{N}{b_1} s \left[ a_1 + s\left( a_3 + a_4 \right) \right]\,, \ \
c_4 = \frac{N}{2b_1} s^2 \left[ a_2 - 2 a_3 \right]\,.
\label{c_sel}
\end{equation}
We note that once again the predictions from the reduced theory match results from simulation extremely well (see \fref{fig_selection_range_parameters}). 

The exponential correction to the result without selection (Eq.(\ref{spdf_mut})) for the well-mixed (single island) case, that is, with $c_4 = 0$, is well-known~\cite{kimura_1964_review,crow_1970,roughgarden_1979,ewens_2004}. Here the metapopulation structure has induced an extra ($c_4 \neq 0$) term, giving a Gaussian form to the correction due to second-order selection effects. One may wonder how such second-order effects can be important. The answer is in some part addressed in~\cite{constable_bio}; in the absence of mutation, if the direction of selection varies from island to island, a migration-selection balance can occur, wiping out first order effects in selection pressure (see parameter $a_{1}$ in \eref{a's}). In these cases the second order term in $s$ can influence the shape of the stationary pdf, as seen in \fref{fig_mig_selec_balance}, where parameters have been chosen which allow this balance. In both plots, the parameters are such that $a_{1}=0$; while the effective theory then predicts that there is no selective pressure at first order, at second order a contribution to the dynamics is made. This second order contribution matches the results from stochastic simulation. Further, as mentioned earlier, at second order the specification of $\sigma_{i}$ and $\rho_{i}$ become important; while both the upper and lower panels of \fref{fig_mig_selec_balance} have the same $\alpha_{i}$ on each island, they have distinct  $\sigma_{i}$ and $\rho_{i}$. This leads to a symmetric stationary pdf in the upper panel, while the lower panel shows the development of an asymmetry. Once again we note that both of these effects are well-predicted by the reduced-dimension theory, \eref{spdf_mut_sel}.

\section{Nature of the fast mode elimination procedure}\label{sec_relate}

The elimination method we use clearly works very well, but it has been presented as a prescription, albeit an intuitively motivated one. This is because the general idea behind the method is very easily understood: at the deterministic level, it is clear that the system should decay to the center manifold or the slow subspace, although the precise definition of the slow subspace may vary. It is also clear how the noise will modify this picture, although once again there are many different ways that this can be implemented mathematically. As a consequence, there is a very large literature on the subject (for an extensive list of references see Ref.~\cite{constable2013}; recent references not given there include \cite{pigolotti2013,rogers_2013,lin_mig_1,lin_mig_2,kogan2014}), and it is variously described as `fast-mode elimination', `adiabatic elimination', `quasi-steady-state approximation', among other terms. While it is an almost impossible task to relate all these methodologies, we will in this section try to give an explanation as to the success of the method we use by applying it to a `toy' model consisting of a linear system with additive noise. We will also explore a different method of reducing the noise which appears more sophisticated, but eventually turns out to perform less well than the method we have used.

By restricting our attention to a system with a linear drift term and a constant diffusion term, the FPE of the system is linear. Further, it is assumed that all the eigenvalues of the system are real and non-positive. If these eigenvalues obey the inequalities
\begin{eqnarray}\label{eigenvalue_ranking}
0 \geq \lambda^{(1)} > \ldots > \lambda^{(r)} \gg  
\lambda^{{(r+1)}} > \ldots > \lambda^{(\mathcal{D})}, 
\end{eqnarray}
then a separation of timescales exists in the system. The fast directions are identified as the right-eigenvectors $\bm{v}^{(r+1)},\ldots,\bm{v}^{(\mathcal{D})}$, while the slow directions are given by the eigenvectors $\bm{v}^{(1)},\ldots,\bm{v}^{(r)}$. These eigenvectors form a basis into which we can transform for clarity. The variables in the slow-fast basis are denoted $\xi_{zi}$ $(i=1,\ldots,r)$ and $\xi_{wj}$ $(j=r+1,\ldots,\mathcal{D})$. In this basis, the dynamics of the system is described by
\begin{eqnarray}
 \frac{d}{dt} \left( \begin{array}{c}  \bm{\xi}_{z} \\ \bm{\xi}_{w} \end{array} \right) = \Lambda \left( \begin{array}{c} \bm{\xi}_{z} \\ \bm{\xi}_{w} \end{array} \right) + \bm{\mu}(t) \,, \label{diagonal_sde}
\end{eqnarray}
where the correlation structure of $\bm{\mu}$ is given by
\begin{eqnarray}
 \langle \mu_{i}(t) \mu_{j}(t') \rangle = B_{ij}\delta(t - t').
\end{eqnarray}
Here the matrix $\Lambda$ is a diagonal matrix of eigenvalues $\lambda^{(i)}$, and the matrix $B$ is constant (independent of the state of the system), since we are assuming the noise to be additive. Finally, if we assume that the boundaries lie at $\pm\infty$, the pdf described by the associated FPE is Gaussian~\cite{lax_1960} (since the FPE is linear) and can thus be described in terms of the time-evolution of its mean and covariance.

The equations for the mean and covariance are soluble~\cite{gardiner_2009}; the solutions for the mean quantities, $\langle\bm{\xi}_{z} \rangle$ and $\langle\bm{\xi}_{w} \rangle$, have the form 
\begin{eqnarray}\label{ODE_general_linear_sol}
\langle \xi_{z i} \rangle &=& c_i e^{\lambda^{(i)} t}\,, \qquad i = 1,\ldots,r \\
\langle \xi_{w j} \rangle &=& c_j e^{\lambda^{(j)} t}\,, \qquad j = r+1,\ldots,\mathcal{D}
\end{eqnarray}
where $c_{i}$ and $c_{j}$ are constants determined from the initial conditions, while the solution for the covariance matrix $\Xi$ is~\cite{van_Kampen_2007}
\begin{eqnarray}
 \Xi(t) = \int_{0}^{t} e^{(t-t')\Lambda }Be^{(t-t') \Lambda} dt' \,. \label{equation_linear_covariance_solution}
\end{eqnarray}
Since in this basis $\Lambda$ is diagonal, the solution for the components of $\Xi$ can be expressed in the particularly neat form 
\begin{eqnarray}
 \Xi_{ij} =  \left(\frac{  e^{(\lambda^{(i)}+\lambda^{(j)}) t} - 1  }{ \lambda^{(i)} + \lambda^{(j)} }\right)  B_{ij} \, , \quad \, i,j=1,\ldots,{\mathcal{D}} \,. \label{full_linear_distribution}
\end{eqnarray}

Having specified the system, we now apply the approximation procedure to it, that is, we  find a reduced form of the SDEs, obtained under the assumption that the inequalities (\ref{eigenvalue_ranking}) hold. To do this, we introduce the partitioned $(r\times r)$ matrix 
\begin{eqnarray}
\bar{\Lambda}_{ij} = \Lambda_{ij} \, , \qquad \forall \, i,j=1,\ldots r \,,
\end{eqnarray}
which is analogous to the $\bar{A}(\bm{z})$ term appearing in \eref{generalSDE1D}, so that the reduced system is  
\begin{eqnarray}
 \frac{d}{dt}\bm{\xi}_{z}  = \bar{\Lambda}  \bm{\xi}_{z} + \bm{\zeta}(t) \,, \label{linear_reduced_sde}
\end{eqnarray}
where $\langle \bm{\zeta}(t) \bm{\zeta}^{T}(t')\rangle = \bar{B}\delta(t-t')$. The structure of $\bar{B}$ is dependent on the projection matrix (see \Aref{app_remove_fast}), which we now discuss.

Constructing the $\mathcal{D} \times r$ matrices $U_{r}$ and $V_{r}$, whose $i^{\rm th}$ columns are defined to be the $i^{\rm th}$ left- and right-eigenvectors, $\bm{u}^{(i)}$ and $\bm{v}^{(i)}$ for $i=1,\ldots r$, the equation for the projection matrix is~\cite{meyer_2000}
\begin{eqnarray}
 P = V_{r} \left( U_{r}^{T}V_{r}\right)^{-1} U_{r}^{T} \,.
\end{eqnarray}
This is a generalization of \eref{defineProjection}, for which the number of slow variables was one, $r=1$. In the particular case of the system (\ref{diagonal_sde}), the projection matrix takes a very simple from as a result of the fact that the system is already in the slow-fast basis. The left- and right-eigenvectors are equal ($\Lambda$ is symmetric) and one finds
\begin{eqnarray}
 P = \left( \begin{array}{cc} I_{r} & 0_{r,m-r} \\ 0_{m-r,r} & 0_{m-r,m-r} \end{array}\right) \, ,
\end{eqnarray}
where $I_{r}$ is the $r\times r$ identity matrix and the $0_{k,l}$ are $k\times l$ zero-matrices. Applying this projection to the noise $\bm{\mu}(t)$, one finds that the form of the correlations of $\bm{\zeta}(t)$ is
\begin{equation}
\bar{B}_{ij} = \left[ P B P^{T} \right]_{ij} \,, \  i,j= 1,\ldots,r.
\label{B_bar_linear_example}
\end{equation}
If we partition the matrix $B$ in the same way as $P$:
\begin{eqnarray}
 B =  \left(\begin{array}{ccc} \mathcal{B}_{11}  & \mathcal{B}_{12} \\  \mathcal{B}_{21} & \mathcal{B}_{22} \end{array}\right),
\label{partition}
\end{eqnarray}
then $\bar{B} = \mathcal{B}_{11}$. 

The solution for the time-evolution of the mean of $\bm{\xi}_{z}$ is clearly unaltered, since the system is linear and $\bar{\Lambda}$ is diagonal and partitioned from $\Lambda$. What form does the the covariance matrix in the reduced system take? Denoting the covariance matrix of the reduced system's pdf $\bar{\Xi}$, from \eref{equation_linear_covariance_solution} we find
\begin{eqnarray}
\bar{\Xi}_{ij}&=&   \left(  \frac{ e^{(\lambda^{(i)}+\lambda^{(j)}) t} - 1  }{ \lambda^{(i)} + \lambda^{(j)} } \right)  \bar{B}_{ij} \, , \nonumber \\
&=&   \left( \frac{  e^{(\lambda^{(i)}+\lambda^{(j)}) t} - 1  }{ \lambda^{(i)} + \lambda^{(j)} } \right) \left[ \mathcal{B}_{11} \right]_{ij}\,, \  i,j=1,\ldots{r} \,.
\label{reduced_linear_covariance_projection}
\end{eqnarray}

Since the system is completely specified by the mean and the variance, we will see that the reduced system is equivalent to taking the limit $\lambda^{(j)}\rightarrow - \infty$ $\forall \, j = r+1 , \ldots \mathcal{D}$ in the original system. In this case, the inequalities (\ref{eigenvalue_ranking}) are enforced to the greatest possible degree. In this limit the elements of the distribution (\ref{full_linear_distribution}) take the form
\begin{eqnarray}
 \Xi_{ij} &\rightarrow&   \left(\frac{  e^{(\lambda^{(i)}+\lambda^{(j)}) t} - 1  }{ \lambda^{(i)} + \lambda^{(j)} } \right) B_{ij} \, , \quad  \,  i,j=1,\ldots,{r} 
\nonumber   \\
 \Xi_{kl} &\rightarrow&  0  \, , \quad    \, k,l={r+1},\ldots ,\mathcal{D}  
\nonumber \\
 \Xi_{il} = \Xi_{li} &\rightarrow&  0  \, , \quad      \,  i=1,\ldots,{r} \quad l={r+1},\ldots ,\mathcal{D} \,.  \label{comparing_limit_lyapunov}
\end{eqnarray}
Comparing this result to \eref{reduced_linear_covariance_projection}, and recalling the partition \eref{partition}, we see that for a system with a linear FPE, the reduced system obtained via the projection matrix method provides an exact agreement with the full system in the limit of $\lambda^{(k)} \rightarrow - \infty$ for $k\geq r + 1$.

We now move to considering how an alternative method, the SDE conditioning method developed in \cite{constable2013}, compares to the projection matrix method when applied to this linear system. The conditioning method bears many similarities to the projection matrix method. As in the projection matrix method, the system is first restricted to a slow subspace, defined by \eref{general_slow_subspace}. Of course, the use of this approximation relies on the identification of a linear slow-fast basis. In the non-linear systems analyzed in \cite{constable2013}, these fast and slow directions were taken to be the eigenvectors of the deterministic system linearized about a fixed point. While formally this results in the deterministic approximation being only locally valid, in the systems addressed in \cite{constable2013} it was found that the approximation remained successful far from the fixed point. However, applying the method to the system described in this section we avoid such concerns, since the system under consideration, \eref{diagonal_sde}, is strictly linear.  

Once the stochastic system is restricted to the slow subspace, there are (by construction) no deterministic dynamics in the fast-variables $\bm{\xi_{w}}$. However there are still stochastic dynamics in the fast-directions which we wish to eliminate. In the method discussed in this paper, the action of the projection matrix on the noise is effectively to neglect the contribution of the noise terms in the fast directions. In \cite{constable2013} however, the restriction of the system to the deterministic slow subspace was made explicit by conditioning the noise matrix on the assumption that there was no noise in the fast-direction. Applying the methodology described there to the system \eref{diagonal_sde}, one once again obtains a reduced system of the form \eref{linear_reduced_sde}, but now the correlation structure of the noise $\bar{B}$ is no longer given by the components of the noise in the slow directions $\mathcal{B}_{11}$. Instead its correlation structure is that of the slow direction conditioned on the noise in the fast direction being zero (see \cite{rogers2012}, Appendix B):
\begin{eqnarray}
 \bar{B} =  \mathcal{B}_{11} - \mathcal{B}_{12} \mathcal{B}_{22}^{-1} \mathcal{B}_{12}^{T} \, .\label{conditioned_B_bar}
\end{eqnarray}
Since the remainder of the reduced equation is identical to the expression arrived at through the projection matrix method, we can simply substitute this into the equation for $\bar{\Xi}$, \eref{reduced_linear_covariance_projection}, to find that the conditioning method predicts
\begin{eqnarray}
 \bar{\Xi}_{ij} = \left(\frac{  e^{(\lambda^{(i)}+\lambda^{(j)}) t} - 1   }{ \lambda^{(i)} + \lambda^{(j)} }\right) \left[ \mathcal{B}_{11} - \mathcal{B}_{12} \mathcal{B}_{22}^{-1} \mathcal{B}_{12}^{T} \right]_{ij} \, , \nonumber \\ \quad \quad \quad  \, i,j=1,\ldots{r} \,. 
\end{eqnarray}
Comparing this with Eqs.~(\ref{reduced_linear_covariance_projection}) and (\ref{comparing_limit_lyapunov}) we see that deterministically both methods reduce to the full system in the limit of $\lambda^{(k)} \rightarrow - \infty$ for $k\geq r + 1$, but while the conditioning method gives a different expression for the noise correlations in the reduced subspace, the projection method gives precisely the right form of the correlations. From this we would expect that the projection matrix method provides a better approximation to the full variable system than conditioning. We would however expect the two methods to give similar results if the correlation between the fast and slow noise variables, $\mathcal{B}_{12}=\mathcal{B}_{21}^{T}$, is small. Numerical investigations show that these statements continue to hold when the techniques are applied to the metapopulation Moran model. The projection matrix method gives a more accurate reduced description of the full system, as the conditioning method tends to underestimate the magnitude of the noise (essentially a consequence of the term $\mathcal{B}_{12} \mathcal{B}_{22}^{-1} \mathcal{B}_{12}^{T}$ in \eref{conditioned_B_bar}). However the results converge as the noise covariance between the fast and slow variables decreases.

In this section it has been shown that in the limit of an infinite timescale separation, the projection matrix method provides an exact description of the time-evolution of the slow variables in a linear FPE. While this only holds for linear systems, it goes some way to explaining the success of the methods in the region of a fixed point (or line of fixed points, in the case of the center manifold). We have not addressed in a formal way however, how the FPE for a full non-linear system may collapse to the results obtained from the projection matrix method. A positive step in this direction would be to relate the method to others described in the literature, for instance the projection operator formalism used in the context of the Fokker-Planck equation~\cite{gardiner_adiabatic_1984}. Nevertheless, given the accuracy of the results we have obtained, we would contend that there may be little need to include any additional contributions from the noise reduction, given the attendant increase in complexity of the formalism that this would entail.

\section{Conclusion}\label{sec_conclusion}

In this paper we have investigated a model of population genetics which included all of the four main evolutionary processes of mutation, genetic drift,  natural selection, and migration between subpopulations, albeit in a simple model of a single locus in haploid individuals having two alleles. However even in this simple setting, it is extremely rare to find studies which include this range of processes, and we are not aware that the FPE containing all of these processes appears in the literature. Of course, part of the reason why the FPE for such a general system has not been constructed is the general belief that it is in any case too complicated to work with, and extracting any meaningful prediction would be extremely difficult, if not impossible. We have shown here that this is not always the case: by using a fast-mode elimination procedure, we have been able to obtain an approximate expression for stationary pdf of this general FPE, which we showed to be in very good agreement with the results of numerical simulations of the original IBM. 

The method of fast-mode elimination (which goes under many other names) is widely used and has an extremely long history. Essentially if there exist modes which decay on time scales which are short compared to those of interest to us, then we expect that we may be able to neglect their detailed dynamics and write down an effective theory in terms of only the `slow modes'. The particular variant that we use here is quite specific: it relies on the system under consideration having a deterministic limit which consists of a set of linear equations together with a small perturbation. Specifically, the perturbations are the processes of mutation and selection which are described in terms are parameters $\kappa_{1i}, \kappa_{2i}$ and $s$, which are sufficiently small that Taylor expansions about the model with no mutation or selection can be truncated at first or second order. The model without mutation or selection is described by the deterministic equations $\dot{x}_i = \sum_{j} H_{ij} x_j$, where $H$ is a constant matrix with a zero eigenvalue and $i, j$ label the islands of the metapopulation. All other eigenvalues of $H$ have a negative real part, and so as long as these are non-zero, one sees immediately that the deterministic system decays to the center manifold defined by the eigenvector of the zero eigenvalue. This picture does not change significantly if stochastic effects (genetic drift) or perturbations (mutation and selection) are added.

Of course, even if this broad picture does not change, it is the effects of genetic drift, mutation and selection that we are interested in, and it is necessary to develop a formalism to be able to calculate these effects. This has been the subject of the current paper in the case when mutation is present. We had previously introduced the method \cite{constable_phys,constable_bio} to deal with genetic drift, but without allowing for mutation. With the inclusion of stochasticity, it was found that an additional condition was needed to ensure the existence of a separation of timescales: the leading non-zero eigenvalue of $H$, $\lambda^{(2)}$, must have a real part whose magnitude is larger than that of the fluctuations, $| \mathrm{Re}(\lambda^{(2)})|> N^{-1/2}$~\cite{constable_phys}, in order to ensure that the deterministic collapse to the center manifold is of sufficient strength to quench fluctuations away from it. Of course, $\lambda^{(2)}$ depends on the network structure, and future work could include trying to understand in more detail for what type of network a timescale separation will be present. For instance, it is clear that if the network consisted of two highly connected clusters, but which were only weakly coupled by migration to one another, the collapse to the center manifold may not occur fast enough for the approximation to be valid. However for a given network and island size structure, we can always calculate the magnitude of the real part of $\lambda^{(2)}$ and check that it is not too small.

Including the process of mutation means that the alleles do not fix, but instead their frequency tends to the stationary pdf previously mentioned. Mutation rates in practice are sufficiently small that their effects need only be treated within a first order calculation. Indeed this assumption, along with that of weak selection, is also essential to the fast-mode elimination procedure described in this paper. The parameters $\kappa_{1 i}, \kappa_{2 i}$ and $s$ must be small in order that the approximations based on the neutral model (namely the existence of fast and slow directions) are not invalidated. Of course, determining precise limits for these parameters is a difficult task, especially for a system as general as that which we present. However, it is clear that the probability of mutation must be much lower than that of migration in order for the approximation to accurately capture the dynamics of the full system. We showed that the effect of mutation and selection to this order in our approximation scheme is equivalent to having a one-island (well-mixed) system with effective mutation rates which depend on the network structure and island size.

Even within the quite specific framework we are working there are variations in the techniques and approaches used. One of these is the exact way that the noise terms in the SDE are treated within the reduction procedure. We specifically examined two different ways of proceeding, and showed why the one we adopted was superior. Our focus has been in obtaining simple effective models from rather complex population genetic models involving generic network structure. Our aim has not been to provide a rigorous mathematical underpinning to the method detailed here, however when applied to the metapopulation Moran model, it is hard to imagine the method performing better. In this sense we have shown that the more complex machinery involved in many fast-variable elimination techniques is not always entirely necessary. To a certain extent, the stochastic nature of the problems investigated contributes to the success of the techniques, as unimportant trends which the reduced model may not capture are averaged out when considering the ensemble. Not least among the strengths of the method is that it can be applied to many other, more complex, problems in population genetics. In fact, most of the restrictions made in this paper can be lifted. For instance, we have already shown elsewhere that the constraint of having fixed island population sizes, inherent in the use of Moran models, can be relaxed~\cite{constable_lv}, and we expect many other generalizations will be possible. We hope to explore and discuss these in future publications.

\begin{appendix}

\section{The diffusion approximation for the metapopulation Moran model}\label{app_diffusion}

In this appendix the diffusion approximation is outlined and applied to the metapopulation Moran model with mutation and selection discussed in the main text. While in principle the dynamics for the process is given by Eqs.~(\ref{master}) and (\ref{eq_rates_mutation_selection}), in practice it is impossible to make analytical progress without making approximations. This usually involves making the diffusion approximation: replacing the discrete variables $n_i$ by $x_i = n_i/\beta_i N$. The assumption is made that $N$ is sufficiently large that $x_i$ can be assumed to be continuous. The form of the transition rates in Eq.~(\ref{eq_rates_mutation_selection}) is such that we can use the same procedure as in the neutral case (discussed in Appendix A of Ref.~\cite{constable_bio}), and so we drop the index on the rates, since the method is the same with and without mutation or selection.

We begin by defining 
\begin{equation}
F^{\pm}_{i}(x_i) \equiv T(N\beta_i x_i \pm 1|N\beta_i x_i)\,,
\label{F_defn}
\end{equation}
so that we may write the master equation (\ref{master}) as
\begin{eqnarray}
\frac{\partial p}{\partial t} &=& \sum_{i=1}^{\mathcal{D}} \left[ F^{+}_{i}(x_{i}- \frac{1}{\beta_{i}N})p(x_{i}-{\frac{1}{\beta_{i}N})} - F^{+}_{i}(x_{i})p(x_{i},t) \right] \nonumber \\
&+&  \sum_{i=1}^{\mathcal{D}} \left[ F^{-}(x_{i}+\frac{1}{\beta_{i}N})p(x_{i}+\frac{1}{\beta_{i}N}) - F^{-}_{i}(x_{i})p(x_{i},t) \right]\,, \nonumber \\
\label{master_F}
\end{eqnarray}
where now $p$ is a function of $\bm{x}=(x_1,\ldots,x_{\mathcal{D}})$ and $t$. The diffusion approximation then consists of performing a Taylor series expansion in $N^{-1}$ (up to order $N^{-2}$) of the terms in the sums in Eq.~(\ref{master_F}), to obtain the FPE (\ref{FPE}) of the main text with~\cite{constable_bio} 
\begin{eqnarray}
A_{i}(\bm{x}) &=& \frac{1}{\beta_i}\,\left[ F^{+}_{i}(\bm{x}) 
- F^{-}_{i}(\bm{x}) \right]\,, \nonumber \\
B_{ii}(\bm{x}) &=& \frac{1}{\beta^2_i}\,\left[ F^{+}_{i}(\bm{x}) 
+ F^{-}_{i}(\bm{x}) \right]. 
\label{AandB}
\end{eqnarray}

As an example, we can use the transition rates for the neutral model given by \eref{eq_neutral_rates} to calculate the $F^{\pm}_{i}(x_i)$. Doing this, and then using Eq.~(\ref{AandB}), gives
\begin{eqnarray}
\label{neutral_A}
A_i(\bm{x}) &=& \sum^{\mathcal{D}}_{j=1} H_{ij} x_{j}, \\
B_{ii}(\bm{x}) &=& \frac{1}{\beta_{i}^{2}}\sum_{j=1}^{\mathcal{D}}G_{ij} \left( x_i + x_j - 2x_i x_j \right),
\label{neutral_B}
\end{eqnarray}
where $G_{ij} = m_{ij} f_j$. The matrix $H_{ij}$, has been introduced since it is central to the understanding of the fast-mode elimination method discussed in \sref{sec_mutation}, and is defined as
\begin{equation}\label{defineH}
 H_{ij}=\frac{G_{ij}}{\beta_{i}} \quad  \, i \neq j, \qquad H_{ii} = -\sum_{j \neq i}^{\mathcal{D}} \frac{G_{ij}}{\beta_{i}} \, .
\end{equation}
The neutral metapopulation Moran model in the diffusion approximation is thus given by the Fokker-Planck equation (FPE) (\ref{FPE}), with a rather simple drift term (Eq.~(\ref{neutral_A})), but a more complex diffusion term (Eq.~(\ref{neutral_B})). The parameters of the model are $m_{ij}, \beta_i$ and $N$; the probability of choosing island $j$ for the birth of an offspring, $f_j$, is taken to be proportional to the population of that island: $f_j = \beta_j/\sum_k \beta_k$, and so is specified by $\bm{\beta}$. Starting from Eqs.~(\ref{eq_rates_selection}) or (\ref{eq_rates_mutation_selection}) for the transition rates with mutation and selection, and proceeding as above, gives us Eqs.~(\ref{A_general}) and (\ref{B_general}) in the main text.

\section{The neutral model: removing fast degrees of freedom}\label{app_remove_fast}

In this appendix the removal of fast degrees of freedom from the neutral metapopulation Moran model is illustrated. This is given by the SDE system (\ref{generalSDE}) with $A_{i}(\bm{x})$ and $B_{ii}(\bm{x})$ given by Eqs.~(\ref{neutral_A}) and (\ref{neutral_B}) respectively. The system in itself is not trivial to solve: the multiplicative nature of the noise makes analytic progress difficult. However the deterministic dynamics are particularly simple. 

The deterministic system is not only entirely linear, but the structure of the matrix $H$, defined in \eref{defineH}, is such that each row sums to exactly zero. This allows one to show~\cite{constable_bio} that the largest eigenvalue of $H$ is zero, $\lambda^{(1)}=0$, and all other eigenvalues have negative real part. The deterministic system therefore quickly collapses along as set of directions specified by the right-eigenvectors $\bm{v}^{(2)},\ldots, \bm{v}^{(\mathcal{D})}$ to lie on a point in the direction $\bm{v}^{(1)}$, along which there are no further dynamics. This subspace is termed the center manifold, and it is this behavior which causes a separation of timescales between the dynamics occurs in the direction $\bm{v}^{(1)}$ and the other directions. While this is not strictly true for the stochastic system, in which $N$ is finite, one would expect that for large enough $N$ this separation of timescales would still be present. This is indeed what has been found~\cite{constable_bio,constable_phys}; the stochastic trajectories quickly collapse along the directions $\bm{v}^{(2)},\ldots, \bm{v}^{(\mathcal{D})}$ to the region of the center manifold. Now however, rather than staying at one point on this line indefinitely, the stochastic dynamics move the system along the line, until fixation of one or other of the alleles occurs. To take advantage of this behavior, we seek to systematically remove the fast-mode components of the system, while leaving the slow mode intact.
 
In the approach to fast-mode elimination that we use in this paper, we assume that there is no noise in the fast direction, so that both $\bm{A}(\bm{x})$ and $\bm{\eta}(\tau)$ are proportional to $\bm{v}^{(1)}$. Since the entries of $\bm{v}^{(1)}$ are all equal (and we take them to be one), the equation of the center manifold is $x_1 =\ldots=x_{\mathcal{D}}$, and we denote this coordinate as $z$. This procedure may be formalized by defining a projection operator~\cite{constable_phys}
\begin{eqnarray}
 P_{ij} = \frac{v^{(1)}_{i}u^{(1)}_{j}}{\sum_{k=1}^{\mathcal{D}} v^{(1)}_{k} u^{(1)}_{k}},
\end{eqnarray}
which when applied to any vector wipes out the fast directions $\bm{v}^{(i)}$ for $i=2, \ldots , \mathcal{D}$, but leaves the component along the direction $\bm{v}^{(1)}$,  untouched. Since $\bm{v}^{(1)}$ has all entries equal to one, and using the orthonormality condition (\ref{orthonormal}), the projection operator reduces to $P_{ij} =  u^{(1)}_{j}$, as given by Eq.~(\ref{defineProjection}) of the main text.

Applying the projection operator to the SDE (\ref{generalSDE}) one finds that
\begin{eqnarray}\label{generalSDE1D}
\dot{z} = \bar{A}(z) + \frac{1}{\sqrt{N}}\zeta(\tau),
\end{eqnarray}
where
\begin{equation}
\bar{A}(z) \equiv \sum^{\mathcal{D}}_{i=1} u^{(1)}_{i} A_{i}(\bm{x}); \ \
\zeta(\tau) \equiv \sum^{\mathcal{D}}_{i=1} u^{(1)}_{i} \eta_{i}(\tau),
\label{def_Abar_zeta}
\end{equation}
and where the bar indicates evaluation on the center manifold. In the case of the neutral metapopulation model, $A_i(\bm{x}) = \sum_j H_{ij}x_j$, and so $\bar{A}=0$, since $\bm{u}^{(1)}$ is a left-eigenvector of $H$ with eigenvalue zero. So the SDE assumes the simple form $\dot{z} = N^{-1/2} \zeta(\tau)$, with $\zeta(\tau)$ a Gaussian correlated white noise with zero mean and correlation function
\begin{eqnarray}
\langle \zeta(\tau) \zeta(\tau') \rangle =  \bar{B}(z)\delta(\tau-\tau'). 
\label{zeta_corr}
\end{eqnarray}
Here 
\begin{equation}
\label{defineBbar_app}
\bar{B}(z) \equiv \sum^{\mathcal{D}}_{i,j=1} u^{(1)}_{i}
\left.B_{ij}(\bm{x})\right|_{\bm{x}=z\bm{v}^{(1)}} u^{(1)}_{j},
\end{equation}
which for the neutral metapopulation model becomes
\begin{equation}
\bar{B}(z) = 2 z(1 - z)\,\sum_{i,k=1}^{\mathcal{D}}\,[u^{(1)}_{i}]^{2}G_{ik}
\beta_{i}^{-2} \equiv 2 b_{1} z (1 - z)\,,
\end{equation}
where we have introduced the constant
\begin{equation}
 b_{1} = \sum_{i,k=1}^{\mathcal{D}}\,[u^{(1)}_{i}]^{2}G_{ik}\beta_{i}^{-2}\,.
\label{b_1}
\end{equation}

Applying this form of fast mode elimination therefore reduces the neutral metapopulation Moran model to an effective well-mixed Moran model of $N$ organisms, but with a noise strength which is increased by the factor of $b_1$ given in Eq.~(\ref{b_1}). Finally, to completely define the reduced model, we need to give the initial value of the system on the center manifold. Since the decay to the center manifold is largely deterministic, we simply take it to be the component of the full initial condition $\bm{x}_{0}$, along $\bm{v}^{1}$:
\begin{eqnarray}
 z_{0} = \sum_{i=1}^{\mathcal{D}}u^{(1)}_{i} x_{0i} \,.
\label{initialConditon}
\end{eqnarray}

The validity of the approximation has been explored in Refs.~\cite{constable_phys} and \cite{constable_bio} by finding the probability of an allele to fixate and the mean time for fixation to occur. These were found by carrying out Gillespie simulations~\cite{gillespie_1976,gillespie_1977} of the original individual based model. They were then compared with the results from an analytic and numerical calculation of the reduced model.

We now turn to the incorporation of other processes to this metapopulation model, such as selection and mutation. The effect of these processes is to break the degeneracy of the deterministic dynamics (see Eqs.~(\ref{A_mutation}) and (\ref{A_general_s})); there is no longer a line of fixed points defined by $x_1 = \ldots = x_{\mathcal{D}}$. This is needed in order to perform a linearization and so define the eigenvectors $\bm{v}^{(i)}$ and $\bm{u}^{(i)}$, which are then used to characterize the fast and slow modes. To combat this, we assume that the nonlinear effects of mutation and selection are small enough that the left- and right-eigenvectors of $H$ remain a good approximation for the fast and slow directions. We may therefore continue to use \eref{general_slow_subspace} to define the slow subspace onto which the system quickly relaxes (but with $A_{i}(\bm{x})$ taken from the system being considered) and the projection matrix $P$ to remove the fast degrees of freedom from the model.

\section{Parameters used in figures}\label{app_parameters}

Throughout this paper, figures which show the stationary pdf of a metapopulation system are accompanied by the stationary pdf of the `well-mixed unstructured analog' of that system. In all these cases, the well-mixed unstructured analog is taken to be a well-mixed system with the same total size as the structured model, and mutation rates corresponding the the average mutation rates across the structured population. Denoting the well mixed parameters with a subscript $wm$, this means the system size of the well-mixed analog may be expressed $N_{wm} = \sum_{i=1}^{ \mathcal{D} }{ \beta_{i} } N  $, while the mutation rates are the weighted means $\kappa_{1wm} = (\sum_{i=1}^{\mathcal{D}} \beta_{i} \kappa_{1i}) / (\sum_{j=1}^{\mathcal{D}} \beta_{j})$ and $\kappa_{2wm} = (\sum_{i=1}^{\mathcal{D}} \beta_{i} \kappa_{2i}) / (\sum_{j=1}^{\mathcal{D}} \beta_{j})$. Likewise, the selection parameters of the well-mixed system are taken to be the weighted means across the structured population; $\sigma_{wm} = (\sum_{i=1}^{\mathcal{D}} \beta_{i} \sigma_{i}) / (\sum_{j=1}^{\mathcal{D}} \beta_{j})$ and $\rho_{wm} = (\sum_{i=1}^{\mathcal{D}} \beta_{i} \rho_{i}) / (\sum_{j=1}^{\mathcal{D}} \beta_{j})$

In \fref{fig_enhance_bistable}, the parameters used in the illustrated system are $\mathcal{D}=2$, $N=300$,
\begin{eqnarray}
m = \left( \begin{array}{cc} 0.7 & 0.04   \\
			     0.3 & 0.96  \end{array}  \right) \, , 
\quad \bm{\beta} = \left( \begin{array}{c} 3.6   \\
					   1.4 \end{array} \right) \, ,
\end{eqnarray}
and
\begin{eqnarray}
 \bm{\kappa}_{1} = \bm{\kappa}_{2} = \left( \begin{array}{c} 1 \times 10^{-5}  \\
							     5 \times 10^{-3}  \end{array}  \right) \, .
\end{eqnarray}
The well-mixed unstructured analog system therefore has $N=1500$ and $\kappa_{1wm}=\kappa_{2wm}\approx1.4\times10^{-3}$.

In \fref{fig_range_parameters} (a), the parameters used in the illustrated system are $\mathcal{D}=6$, $N=300$, $m_{ii}=0.85$, $m_{ij} = 0.03 \, (i\neq j)$,
\begin{eqnarray}
\bm{\beta} = \left( \begin{array}{c}       4   \\
					   1   \\
                                           1   \\
                                           2   \\
                                           1   \\
                                           3 \end{array} \right) \, ,
\end{eqnarray}
and
\begin{eqnarray}
\bm{\kappa}_{1} = \left( \begin{array}{c}  1\times10^{-4}   \\
					   7.5\times10^{-5} \\
                                           0                \\
                                           2.5\times10^{-4} \\
                                           2.5\times10^{-4} \\
					   0  \end{array}  \right) \, ,
\quad \quad
\bm{\kappa}_{2} = \left( \begin{array}{c}  6\times10^{-4}   \\
					   2\times10^{-4} \\
                                           0                \\
                                           2\times10^{-4} \\
                                           2\times10^{-4} \\
					   1\times10^{-4}  \end{array}  \right) \, .
\end{eqnarray}
The well-mixed unstructured analog system therefore has $N=3600$, $\kappa_{1wm}=1.02\times10^{-4}$ and $\kappa_{2wm}=2.92\times10^{-4}$.

In \fref{fig_range_parameters} (b), the parameters used in the illustrated system are $\mathcal{D}=8$, $N=200$, $m_{ii}=0.8$, $m_{ij} = 0.2/7 \, (i\neq j)$,
\begin{eqnarray}
\bm{\beta} = \left( \begin{array}{c} 	   1   \\
					   1   \\
                                           0.4   \\
                                           2   \\
                                           1   \\
                                           3   \\
                                           2   \\
                                           1 \end{array} \right) \, ,
\end{eqnarray}
and
\begin{eqnarray}
\bm{\kappa}_{1} = \left( \begin{array}{c}  4\times10^{-3} \\
					   3\times10^{-3} \\
                                           0              \\
                                           1\times10^{-3} \\
                                           1\times10^{-3} \\
					   0              \\
                                           1\times10^{-3} \\ 
				           3\times10^{-4} \\  \end{array}  \right) \, ,
\quad \quad
\bm{\kappa}_{2} = \left( \begin{array}{c}  0              \\
					   3\times10^{-3} \\
                                           1\times10^{-3} \\
                                           0               \\
                                           1\times10^{-3} \\
					   1\times10^{-3} \\
                                           5\times10^{-4} \\ 
				           1\times10^{-3} \\  \end{array}  \right) \, .
\end{eqnarray}
The well-mixed unstructured analog system therefore has $N=2280$ and $\kappa_{1wm}=1.08\times10^{-3}$ and $\kappa_{2wm}=8.25\times10^{-4}$.

In \fref{fig_range_parameters} (c), the parameters used in the illustrated system are $\mathcal{D}=4$, $N=200$,
\begin{eqnarray}
m = \left( \begin{array}{cccc} 0.9 & 0.05 & 0.03 & 0.01 \\ 
			     0.025 & 0.85 & 0.04 & 0.04 \\ 
			     0.05 & 0.05 & 0.88 & 0.03 \\ 
			     0.025 & 0.05 & 0.05 & 0.92  \end{array}  \right) \, , 
\quad \bm{\beta} = \left( \begin{array}{c} 1   \\
                                           1   \\
                                           1   \\
					   1 \end{array} \right) \, ,
\end{eqnarray}
and
\begin{eqnarray}
\bm{\kappa}_{1} = \left( \begin{array}{c}  4\times10^{-3} \\
					   3\times10^{-3} \\
                                           0             \\ 
				           1\times10^{-3}   \end{array}  \right) \, ,
\quad \quad
\bm{\kappa}_{2} = \left( \begin{array}{c}  0 \\
					   3\times10^{-3} \\
                                           1\times10^{-3} \\
				           0                 \end{array}  \right) \,. 
\end{eqnarray}
The well-mixed unstructured analog system therefore has $N=800$, $\kappa_{1wm}=2\times10^{-3}$ and $\kappa_{2wm}=1\times10^{-3}$.

In \fref{fig_range_parameters} (d), the parameters used in the illustrated system are $\mathcal{D}=5$, $N=200$,
\begin{eqnarray}
m = \left( \begin{array}{ccccc} 0.9 & 0.05 & 0.0 & 0.01 & 0 \\
                                0.025 & 0.85 & 0.02 & 0.04 & 0.025 \\
                                0.05 & 0.025 & 0.88 & 0.025 & 0.025 \\
                                0.025 & 0.025 & 0.05 & 0.92 & 0.05 \\
                                0 & 0.05 & 0.05 & 0.005 & 0.9 \end{array}  \right) \, , 
\quad \bm{\beta} = \left( \begin{array}{c} 1   \\
                                           1   \\
                                           1   \\
                                           1   \\
					   1 \end{array} \right) \, ,
\end{eqnarray}
and
\begin{eqnarray}
\bm{\kappa}_{1} = \left( \begin{array}{c}  0              \\
					   3\times10^{-3} \\
                                           0             \\ 
				           1\times10^{-3}\\
				           2\times10^{-3}  \end{array}  \right) \, ,
\quad \quad
\bm{\kappa}_{2} = \left( \begin{array}{c}  0              \\
					   4\times10^{-3} \\
                                           1\times10^{-3} \\
				           0              \\  
                                           2\times10^{-3} \end{array}  \right) \,.
\end{eqnarray}
The well-mixed unstructured analog system is therefore taken to have $N=1000$, $\kappa_{1wm}=1.2\times10^{-3}$ and $\kappa_{2wm}=1.4\times10^{-3}$.

In \fref{fig_selection_range_parameters} (a), the parameters used are $\mathcal{D}=3$, $N=230$, $m_{ii}=0.9$, $m_{ij} = 0.05 \, (i\neq j)$, $s=2.15\times10^{-3}$,
\begin{eqnarray}
\bm{\beta} = \left( \begin{array}{c}       1   \\
					   2   \\
                                           4    \end{array} \right) \, ,
\end{eqnarray}
and $\kappa_{1i}=2\times10^{-4}$, $\kappa_{2i}=1\times10^{-4}$,
\begin{eqnarray}
\bm{\rho} = \left( \begin{array}{c}  0   \\
				      0\\
                                       1   \end{array}  \right) \, ,
\quad \quad
\bm{\sigma} = \left( \begin{array}{c}  1  \\
				       1 \\
                                       0  \end{array}  \right) \, .
\end{eqnarray}
The well-mixed unstructured analog system therefore has $N=1610$, $\kappa_{1wm}=2\times10^{-4}$, $\kappa_{2wm}=1\times10^{-4}$, $\sigma_{wm}=3/7$ and $\rho_{wm}=4/7$.

In \fref{fig_selection_range_parameters} (b), the parameters used are $\mathcal{D}=4$, $N=400$, $m_{ii}=0.8$, $m_{ij} = 0.2/3 \, (i\neq j)$, $s=2.5\times10^{-3}$,
\begin{eqnarray}
\bm{\beta} = \left( \begin{array}{c}       1   \\
					   1   \\
					   2   \\
                                           1    \end{array} \right) \, ,
\end{eqnarray}
and
\begin{eqnarray}
\bm{\kappa}_{1} = \left( \begin{array}{c}  6\times10^{-4} \\
					   6\times10^{-4} \\
                                           6\times10^{-4}\\ 
				           6\times10^{-4}\end{array}  \right) \, ,
\quad \quad
\bm{\kappa}_{2} = \left( \begin{array}{c}  2\times10^{-3} \\
					   4\times10^{-3} \\
                                           1\times10^{-4} \\
				           6\times10^{-4} \end{array}  \right) \, 
\end{eqnarray}
with
\begin{eqnarray}
\bm{\rho} = \left( \begin{array}{c}  1   \\
				     0   \\
                                     1   \\
				     0  \end{array}  \right) \, ,
\quad \quad
\bm{\sigma} = \left( \begin{array}{c}  -1  \\
				       2 \\
                                       0 \\
				      1  \end{array}  \right) \, .
\end{eqnarray}
The well-mixed unstructured analog system therefore has $N=2000$, $\kappa_{1wm}=6\times10^{-4}$, $\kappa_{2wm}=1.36\times10^{-3}$, $\sigma_{wm}=2/5$ and $\rho_{wm}=3/5$.

In \fref{fig_selection_range_parameters} (c), the parameters used are $\mathcal{D}=5$, $N=200$, $m_{ii}=0.85$, $m_{ij} = 3.75\times10^{2} \, (i\neq j)$, $s=1.5\times10^{-2}$,
\begin{eqnarray}
\bm{\beta} = \left( \begin{array}{c}       2   \\
					   1   \\
					   2   \\
                                           1    \\
					   2\end{array} \right) \, ,
\end{eqnarray}
and $\kappa_{1i}=1\times10^{-4}=\kappa_{2i}=1\times10^{-4}$,
\begin{eqnarray}
\bm{\rho} = \left( \begin{array}{c}  1   \\
				     1   \\
                                     1   \\
				     0  \\
			             0   \end{array}  \right) \, ,
\quad \quad
\bm{\sigma} = \left( \begin{array}{c}  0  \\
				       0 \\
                                       0 \\
				       0 \\
					1  \end{array}  \right) \, .
\end{eqnarray}
The well-mixed unstructured analog system therefore has $N=1600$, $\kappa_{1wm}=1\times10^{-4}$, $\kappa_{2wm}=1\times10^{-4}$, $\sigma_{wm}=1/4$ and $\rho_{wm}=5/8$.

In \fref{fig_selection_range_parameters} (d), the parameters used are $\mathcal{D}=6$, $N=400$, $m_{ii}=0.8$, $m_{ij} = 4\times10^{2} \, (i\neq j)$, $s=5\times10^{-3}$,
\begin{eqnarray}
\bm{\beta} = \left( \begin{array}{c}       1   \\
					   1   \\
					   2   \\
                                           1    \\
					   3    \\
                                           1    \end{array} \right) \, ,
\end{eqnarray}
and
\begin{eqnarray}
\bm{\kappa}_{1} = \left( \begin{array}{c}  6\times10^{-4} \\
					   6\times10^{-4} \\
                                           6\times10^{-4}\\ 
				           6\times10^{-3}\\
					   0            \\
					    0\end{array}  \right) \, ,
\quad \quad
\bm{\kappa}_{2} = \left( \begin{array}{c}  2\times10^{-3} \\
					   4\times10^{-3} \\
                                           1\times10^{-3} \\
				           6\times10^{-4} \\
                                           0               \\
                                           0 \end{array}  \right) \, 
\end{eqnarray}
with
\begin{eqnarray}
\bm{\rho} = \left( \begin{array}{c}  1   \\
				     0   \\
                                     1   \\
				     0  \\
			             1 \\
				     -1 \end{array}  \right) \, ,
\quad \quad
\bm{\sigma} = \left( \begin{array}{c}  -1  \\
				       2 \\
                                       0 \\
				       1 \\
					1 \\
                                        -1  \end{array}  \right) \, .
\end{eqnarray}
The well-mixed unstructured analog system therefore has $N=3600$, $\kappa_{1wm}=9.3\times10^{-4}$, $\kappa_{2wm}=9.6\times10^{-4}$, $\sigma_{wm}=4/9$ and $\rho_{wm}=5/9$.

In the upper panel of \fref{fig_mig_selec_balance}, the parameters used are are $\mathcal{D}=2$, $N=400$, $m_{ii}=0.85$, $m_{ij} = 0.15 \, (i\neq j)$, $s=1\times10^{-2}$, $\bm{\beta}=(1,1)$,
and $\kappa_{1i}=\kappa_{2i}=6\times10^{-4}$,
\begin{eqnarray}
\bm{\rho} = \left( \begin{array}{c}  1   \\
                                       -1   \end{array}  \right) \, ,
\quad \quad
\bm{\sigma} = \left( \begin{array}{c}  -1  \\
				       1   \end{array}  \right) \, .
\end{eqnarray}
The well-mixed unstructured analog system therefore has $N=800$, $\kappa_{1wm}=\kappa_{2wm}=6\times10^{-4}$, $\sigma_{wm}=\rho_{wm}=0$.

In the lower panel of \fref{fig_mig_selec_balance}, the parameters used are are $\mathcal{D}=2$, $N=400$, $m_{ii}=0.8$, $m_{ij} = 0.2 \, (i\neq j)$, $s=1.5\times10^{-2}$, $\bm{\beta}=(1,1)$,
and $\kappa_{1i}=\kappa_{2i}=6\times10^{-4}$,
\begin{eqnarray}
\bm{\rho} = \left( \begin{array}{c}  1   \\
                                     1   \end{array}  \right) \, ,
\quad \quad
\bm{\sigma} = \left( \begin{array}{c}  -1  \\
				       3   \end{array}  \right) \, .
\end{eqnarray}
The well-mixed unstructured analog system therefore has $N=800$, $\kappa_{1wm}=\kappa_{2wm}=6\times10^{-4}$, $\sigma_{wm}=\rho_{wm}=1$.

%\section{Fitness normalisation}\label{app_selection_normalisation}

\end{appendix}

%\bibliography{mutations2}

\begin{thebibliography}{31}
\expandafter\ifx\csname natexlab\endcsname\relax\def\natexlab#1{#1}\fi
\expandafter\ifx\csname bibnamefont\endcsname\relax
  \def\bibnamefont#1{#1}\fi
\expandafter\ifx\csname bibfnamefont\endcsname\relax
  \def\bibfnamefont#1{#1}\fi
\expandafter\ifx\csname citenamefont\endcsname\relax
  \def\citenamefont#1{#1}\fi
\expandafter\ifx\csname url\endcsname\relax
  \def\url#1{\texttt{#1}}\fi
\expandafter\ifx\csname urlprefix\endcsname\relax\def\urlprefix{URL }\fi
\providecommand{\bibinfo}[2]{#2}
\providecommand{\eprint}[2][]{\url{#2}}

\bibitem[{\citenamefont{Roughgarden}(1979)}]{roughgarden_1979}
\bibinfo{author}{\bibfnamefont{J.}~\bibnamefont{Roughgarden}},
  \emph{\bibinfo{title}{{T}heory of {P}opulation {G}enetics and {E}volutionary
  {E}cology: an {I}ntroduction}} (\bibinfo{publisher}{Macmillan},
  \bibinfo{address}{New York}, \bibinfo{year}{1979}).

\bibitem[{\citenamefont{Halliburton}(2004)}]{halliburton_2004}
\bibinfo{author}{\bibfnamefont{R.}~\bibnamefont{Halliburton}},
  \emph{\bibinfo{title}{{I}ntroduction to {P}opulation {G}enetics}}
  (\bibinfo{publisher}{Pearson Press}, \bibinfo{address}{New Jersey},
  \bibinfo{year}{2004}).

\bibitem[{\citenamefont{Hartl and Clark}(2007)}]{hartl_1989}
\bibinfo{author}{\bibfnamefont{D.~L.} \bibnamefont{Hartl}} \bibnamefont{and}
  \bibinfo{author}{\bibfnamefont{A.~G.} \bibnamefont{Clark}},
  \emph{\bibinfo{title}{Principles of Population Genetics}}
  (\bibinfo{publisher}{Sinauer Associates Inc.}, \bibinfo{address}{Sunderland,
  Mass.}, \bibinfo{year}{2007}), \bibinfo{note}{{Fourth edition}}.

\bibitem[{\citenamefont{Constable and
  McKane}(2014{\natexlab{a}})}]{constable_phys}
\bibinfo{author}{\bibfnamefont{G.~W.~A.} \bibnamefont{Constable}}
  \bibnamefont{and} \bibinfo{author}{\bibfnamefont{A.~J.}
  \bibnamefont{McKane}}, \bibinfo{journal}{Phys. Rev. E}
  \textbf{\bibinfo{volume}{89}}, \bibinfo{pages}{032141}
  (\bibinfo{year}{2014}{\natexlab{a}}).

\bibitem[{\citenamefont{Constable and
  McKane}(2014{\natexlab{b}})}]{constable_bio}
\bibinfo{author}{\bibfnamefont{G.~W.~A.} \bibnamefont{Constable}}
  \bibnamefont{and} \bibinfo{author}{\bibfnamefont{A.~J.}
  \bibnamefont{McKane}}, \bibinfo{journal}{J. Theor. Biol.}
  \textbf{\bibinfo{volume}{358}}, \bibinfo{pages}{149}
  (\bibinfo{year}{2014}{\natexlab{b}}).

\bibitem[{\citenamefont{Constable et~al.}(2013)\citenamefont{Constable, McKane,
  and Rogers}}]{constable2013}
\bibinfo{author}{\bibfnamefont{G.~W.~A.} \bibnamefont{Constable}},
  \bibinfo{author}{\bibfnamefont{A.~J.} \bibnamefont{McKane}},
  \bibnamefont{and} \bibinfo{author}{\bibfnamefont{T.}~\bibnamefont{Rogers}},
  \bibinfo{journal}{J. Phys. A: Math. Theor.} \textbf{\bibinfo{volume}{46}},
  \bibinfo{pages}{295002} (\bibinfo{year}{2013}).

\bibitem[{\citenamefont{Ewens}(2004)}]{ewens_2004}
\bibinfo{author}{\bibfnamefont{W.~J.} \bibnamefont{Ewens}},
  \emph{\bibinfo{title}{Mathematical Population Genetics}}
  (\bibinfo{publisher}{Springer-Verlag}, \bibinfo{address}{Berlin},
  \bibinfo{year}{2004}), \bibinfo{note}{{Second edition}}.

\bibitem[{\citenamefont{van Kampen}(2007)}]{van_Kampen_2007}
\bibinfo{author}{\bibfnamefont{N.~G.} \bibnamefont{van Kampen}},
  \emph{\bibinfo{title}{Stochastic {P}rocesses in {P}hysics and {C}hemistry}}
  (\bibinfo{publisher}{Elsevier}, \bibinfo{address}{Amsterdam},
  \bibinfo{year}{2007}).

\bibitem[{\citenamefont{Moran}(1962)}]{moran_1962}
\bibinfo{author}{\bibfnamefont{P.~A.~P.} \bibnamefont{Moran}},
  \emph{\bibinfo{title}{{T}he {S}tatistical {P}rocesses of {E}volutionary
  {T}heory}} (\bibinfo{publisher}{Clarendon Press}, \bibinfo{address}{Oxford},
  \bibinfo{year}{1962}).

\bibitem[{\citenamefont{Blythe and McKane}(2007)}]{blythe_mckane_models_2007}
\bibinfo{author}{\bibfnamefont{R.~A.} \bibnamefont{Blythe}} \bibnamefont{and}
  \bibinfo{author}{\bibfnamefont{A.~J.} \bibnamefont{McKane}},
  \bibinfo{journal}{J. Stat. Mech.} \textbf{\bibinfo{volume}{P07018}}
  (\bibinfo{year}{2007}).

\bibitem[{\citenamefont{Rohner et~al.}(2013)\citenamefont{Rohner, Jarosz,
  Kowalko, Yoshizawa, Jeffery, Borowsky, Lindquist, and Tabin}}]{mexican_fish}
\bibinfo{author}{\bibfnamefont{N.}~\bibnamefont{Rohner}},
  \bibinfo{author}{\bibfnamefont{D.~F.} \bibnamefont{Jarosz}},
  \bibinfo{author}{\bibfnamefont{J.~E.} \bibnamefont{Kowalko}},
  \bibinfo{author}{\bibfnamefont{M.}~\bibnamefont{Yoshizawa}},
  \bibinfo{author}{\bibfnamefont{W.~R.} \bibnamefont{Jeffery}},
  \bibinfo{author}{\bibfnamefont{R.~L.} \bibnamefont{Borowsky}},
  \bibinfo{author}{\bibfnamefont{S.}~\bibnamefont{Lindquist}},
  \bibnamefont{and} \bibinfo{author}{\bibfnamefont{C.~J.} \bibnamefont{Tabin}},
  \bibinfo{journal}{Science} \textbf{\bibinfo{volume}{342}},
  \bibinfo{pages}{1372} (\bibinfo{year}{2013}).

\bibitem[{\citenamefont{Gardiner}(2009)}]{gardiner_2009}
\bibinfo{author}{\bibfnamefont{C.~W.} \bibnamefont{Gardiner}},
  \emph{\bibinfo{title}{{H}andbook of {S}tochastic {M}ethods}}
  (\bibinfo{publisher}{Springer}, \bibinfo{address}{Berlin},
  \bibinfo{year}{2009}), \bibinfo{edition}{4th} ed.

\bibitem[{\citenamefont{Risken}(1989)}]{risken_1989}
\bibinfo{author}{\bibfnamefont{H.}~\bibnamefont{Risken}},
  \emph{\bibinfo{title}{The {F}okker-{P}lanck {E}quation}}
  (\bibinfo{publisher}{Springer}, \bibinfo{address}{Berlin},
  \bibinfo{year}{1989}), \bibinfo{edition}{2nd} ed.

\bibitem[{\citenamefont{Crow and Kimura}(1970)}]{crow_1970}
\bibinfo{author}{\bibfnamefont{J.~F.} \bibnamefont{Crow}} \bibnamefont{and}
  \bibinfo{author}{\bibfnamefont{M.}~\bibnamefont{Kimura}},
  \emph{\bibinfo{title}{An {I}ntroduction to {P}opulation {G}enetics {T}heory}}
  (\bibinfo{publisher}{Harper and Row}, \bibinfo{address}{New York},
  \bibinfo{year}{1970}).

\bibitem[{\citenamefont{Maruyama}(1977)}]{maruyama_1977}
\bibinfo{author}{\bibfnamefont{T.}~\bibnamefont{Maruyama}},
  \emph{\bibinfo{title}{{S}tochastic {P}roblems in {P}opulation {G}enetics}}
  (\bibinfo{publisher}{Springer}, \bibinfo{address}{Berlin},
  \bibinfo{year}{1977}).

\bibitem[{\citenamefont{Lynch}(2010)}]{lynch2010}
\bibinfo{author}{\bibfnamefont{M.}~\bibnamefont{Lynch}},
  \bibinfo{journal}{Trends Genet.} \textbf{\bibinfo{volume}{26}},
  \bibinfo{pages}{345} (\bibinfo{year}{2010}).

\bibitem[{\citenamefont{Nielsen and Yang}(2003)}]{nielson2003}
\bibinfo{author}{\bibfnamefont{R.}~\bibnamefont{Nielsen}} \bibnamefont{and}
  \bibinfo{author}{\bibfnamefont{Z.}~\bibnamefont{Yang}},
  \bibinfo{journal}{Mol. Biol. Evol.} \textbf{\bibinfo{volume}{20}},
  \bibinfo{pages}{1231} (\bibinfo{year}{2003}).

\bibitem[{\citenamefont{Barrett et~al.}(2006)\citenamefont{Barrett, {L. K.
  M'Gonigle}, and Otto}}]{barret2006}
\bibinfo{author}{\bibfnamefont{R.~D.~H.} \bibnamefont{Barrett}},
  \bibinfo{author}{\bibnamefont{{L. K. M'Gonigle}}}, \bibnamefont{and}
  \bibinfo{author}{\bibfnamefont{S.~P.} \bibnamefont{Otto}},
  \bibinfo{journal}{Genetics} \textbf{\bibinfo{volume}{174}},
  \bibinfo{pages}{2071} (\bibinfo{year}{2006}).

\bibitem[{\citenamefont{Kimura}(1964)}]{kimura_1964_review}
\bibinfo{author}{\bibfnamefont{M.}~\bibnamefont{Kimura}}, \bibinfo{journal}{J.
  Appl. Probab.} \textbf{\bibinfo{volume}{1}}, \bibinfo{pages}{177}
  (\bibinfo{year}{1964}).

\bibitem[{\citenamefont{Pigolotti et~al.}(2013)\citenamefont{Pigolotti, Benzi,
  Perlekar, Jensen, Toschi, and Nelson}}]{pigolotti2013}
\bibinfo{author}{\bibfnamefont{S.}~\bibnamefont{Pigolotti}},
  \bibinfo{author}{\bibfnamefont{R.}~\bibnamefont{Benzi}},
  \bibinfo{author}{\bibfnamefont{P.}~\bibnamefont{Perlekar}},
  \bibinfo{author}{\bibfnamefont{H.~H.} \bibnamefont{Jensen}},
  \bibinfo{author}{\bibfnamefont{F.}~\bibnamefont{Toschi}}, \bibnamefont{and}
  \bibinfo{author}{\bibfnamefont{D.~R.} \bibnamefont{Nelson}},
  \bibinfo{journal}{Theo. Popul. Biol.} \textbf{\bibinfo{volume}{84}},
  \bibinfo{pages}{72} (\bibinfo{year}{2013}).

\bibitem[{\citenamefont{Rogers and Gross}(2013)}]{rogers_2013}
\bibinfo{author}{\bibfnamefont{T.}~\bibnamefont{Rogers}} \bibnamefont{and}
  \bibinfo{author}{\bibfnamefont{T.}~\bibnamefont{Gross}},
  \bibinfo{journal}{Phys. Rev. E} \textbf{\bibinfo{volume}{88}},
  \bibinfo{pages}{030102(R)} (\bibinfo{year}{2013}).

\bibitem[{\citenamefont{Lin et~al.}(2015{\natexlab{a}})\citenamefont{Lin, Kim,
  and Doering}}]{lin_mig_1}
\bibinfo{author}{\bibfnamefont{Y.~T.} \bibnamefont{Lin}},
  \bibinfo{author}{\bibfnamefont{H.}~\bibnamefont{Kim}}, \bibnamefont{and}
  \bibinfo{author}{\bibfnamefont{C.~R.} \bibnamefont{Doering}},
  \bibinfo{journal}{J. Math. Biol.} \textbf{\bibinfo{volume}{70}},
  \bibinfo{pages}{647} (\bibinfo{year}{2015}{\natexlab{a}}).

\bibitem[{\citenamefont{Lin et~al.}(2015{\natexlab{b}})\citenamefont{Lin, Kim,
  and Doering}}]{lin_mig_2}
\bibinfo{author}{\bibfnamefont{Y.~T.} \bibnamefont{Lin}},
  \bibinfo{author}{\bibfnamefont{H.}~\bibnamefont{Kim}}, \bibnamefont{and}
  \bibinfo{author}{\bibfnamefont{C.~R.} \bibnamefont{Doering}},
  \bibinfo{journal}{J. Math. Biol.} \textbf{\bibinfo{volume}{70}},
  \bibinfo{pages}{679} (\bibinfo{year}{2015}{\natexlab{b}}).

\bibitem[{\citenamefont{Kogan et~al.}(2014)\citenamefont{Kogan, Khasin,
  Meerson, Schneider, and Myers}}]{kogan2014}
\bibinfo{author}{\bibfnamefont{O.}~\bibnamefont{Kogan}},
  \bibinfo{author}{\bibfnamefont{M.}~\bibnamefont{Khasin}},
  \bibinfo{author}{\bibfnamefont{B.}~\bibnamefont{Meerson}},
  \bibinfo{author}{\bibfnamefont{D.}~\bibnamefont{Schneider}},
  \bibnamefont{and} \bibinfo{author}{\bibfnamefont{C.~R.} \bibnamefont{Myers}},
  \bibinfo{journal}{Phys. Rev. E} \textbf{\bibinfo{volume}{90}},
  \bibinfo{pages}{042149} (\bibinfo{year}{2014}).

\bibitem[{\citenamefont{Lax}(1960)}]{lax_1960}
\bibinfo{author}{\bibfnamefont{M.}~\bibnamefont{Lax}}, \bibinfo{journal}{Rev.
  Mod. Phys.} \textbf{\bibinfo{volume}{32}}, \bibinfo{pages}{25}
  (\bibinfo{year}{1960}).

\bibitem[{\citenamefont{Meyer}(2000)}]{meyer_2000}
\bibinfo{author}{\bibfnamefont{C.~D.} \bibnamefont{Meyer}},
  \emph{\bibinfo{title}{{M}atrix {A}nalysis and {A}pplied {Linear} {A}lgebra}}
  (\bibinfo{publisher}{SIAM}, \bibinfo{address}{Philadelphia},
  \bibinfo{year}{2000}).

\bibitem[{\citenamefont{Rogers et~al.}(2012)\citenamefont{Rogers, McKane, and
  Rossberg}}]{rogers2012}
\bibinfo{author}{\bibfnamefont{T.}~\bibnamefont{Rogers}},
  \bibinfo{author}{\bibfnamefont{A.~J.} \bibnamefont{McKane}},
  \bibnamefont{and} \bibinfo{author}{\bibfnamefont{A.~G.}
  \bibnamefont{Rossberg}}, \bibinfo{journal}{Phys. Biol.}
  \textbf{\bibinfo{volume}{9}}, \bibinfo{pages}{066002} (\bibinfo{year}{2012}).

\bibitem[{\citenamefont{Gardiner}(1984)}]{gardiner_adiabatic_1984}
\bibinfo{author}{\bibfnamefont{C.~W.} \bibnamefont{Gardiner}},
  \bibinfo{journal}{Phys. Rev. A} \textbf{\bibinfo{volume}{29}},
  \bibinfo{pages}{2814} (\bibinfo{year}{1984}).

\bibitem[{\citenamefont{Constable and McKane}(2015)}]{constable_lv}
\bibinfo{author}{\bibfnamefont{G.~W.~A.} \bibnamefont{Constable}}
  \bibnamefont{and} \bibinfo{author}{\bibfnamefont{A.~J.}
  \bibnamefont{McKane}}, \bibinfo{journal}{Phys. Rev. Lett.}
  \textbf{\bibinfo{volume}{114}}, \bibinfo{pages}{038101}
  (\bibinfo{year}{2015}).

\bibitem[{\citenamefont{Gillespie}(1976)}]{gillespie_1976}
\bibinfo{author}{\bibfnamefont{D.~T.} \bibnamefont{Gillespie}},
  \bibinfo{journal}{J. Comput. Phys.} \textbf{\bibinfo{volume}{22}},
  \bibinfo{pages}{403} (\bibinfo{year}{1976}).

\bibitem[{\citenamefont{Gillespie}(1977)}]{gillespie_1977}
\bibinfo{author}{\bibfnamefont{D.~T.} \bibnamefont{Gillespie}},
  \bibinfo{journal}{J. Phys. Chem.} \textbf{\bibinfo{volume}{81}},
  \bibinfo{pages}{2340} (\bibinfo{year}{1977}).

\end{thebibliography}

\end{document}